\documentclass[10pt]{iopart}

\usepackage{iopams}
\usepackage[T1]{fontenc}
\usepackage[utf8]{inputenc}
\usepackage{textcomp}
\usepackage{float}
\usepackage{graphicx}
\usepackage{tabularx}
\usepackage{subfigure}
\usepackage{grffile}
\usepackage{caption}
\usepackage[dvipsnames]{xcolor}
\usepackage{ulem}
\usepackage{soul}
\usepackage{enumerate}
\usepackage{hyperref}
\usepackage{url}
\usepackage{cite}
\usepackage{comment}
\usepackage{todonotes}
\usepackage{multirow}
\usepackage{booktabs}

\graphicspath{{./Figures/}}

\newcommand{\rev}[1]{{{#1}}}

\setulcolor{blue}
\setstcolor{red}

\soulregister\cite7
\soulregister\ref7
\soulregister\eref7
\soulregister\Eref7
\soulregister\fref7
\soulregister\Fref7
\soulregister\sref7
\soulregister\Sref7
\soulregister\tref7
\soulregister\Tref7
\soulregister\pageref7

\begin{document}

\title{\rev{A computational study on the energy efficiency of species production by single-pulse streamers in air}}

\author{Baohong Guo, Jannis Teunissen$^{*}$}

\address{Centrum Wiskunde \& Informatica (CWI), Amsterdam, The Netherlands}

\ead{jannis.teunissen@cwi.nl}

\vspace{10pt}

\begin{indented}
\item[]
\today
\end{indented}

\begin{abstract}
  We study the energy efficiency of species production by streamer discharges with a single voltage pulse in atmospheric dry air, using a 2D axisymmetric fluid model.
  Sixty different positive streamers are simulated by varying the electrode \rev{geometry}, the pulse duration and the applied voltage.
  Between these cases, the streamer radius and velocity vary by about an order of magnitude, but the variation in the maximal electric field is significantly smaller, about 30\%.
  We find that $G$-values for the production of N($^4$S), O($^3$P), NO and N$_2$O, which have relatively high activation energies, vary by about 30\% to \rev{60\%}.
  This variation is mainly caused by two factors: differences in the fraction of energy deposited in the streamer head region, and differences in the maximal electric field at the streamer head.
  When accounting for both factors, our computed $G$-values are in good agreement with an analytic estimate proposed by Naidis (2012 \textit{Plasma Sources Sci.~Technol.}~\textbf{21} 042001).
  We also simulate negative streamers and find that their production of N($^4$S), O($^3$P) and NO is less energy efficient.
  The results suggest that energy efficiency can be increased by reducing Joule heating in the streamer channel and by increasing the maximal electric field at the streamer head, for example by using short voltage pulses with a high applied voltage.
\end{abstract}

%
%
%
\ioptwocol

\section{Introduction}\label{sec:introduction}

Streamer discharges are fast-moving ionization fronts with self-organized field enhancement at their heads~\cite{nijdam2020a}. 
In such discharges, chemically active species are produced by collisions of energetic electrons with gas molecules~\cite{wang2020a}.
Due to their highly non-equilibrium nature~\cite{bruggeman2017}, streamer discharges can efficiently produce species with a high activation energy without significant gas heating.
Streamer discharges are used for many plasma chemical applications such as air purification or ozone production~\cite{kim2004, winands2006, he2019, jiang2022}, removal of nitrogen oxides~\cite{amirov1998, hazama2000, kuroki2002}, liquid treatment~\cite{sahni2006, joshi2013}, surface modification~\cite{bardos2010, polonskyi2021}, plasma medicine~\cite{fridman2008, vonwoedtke2020} and plasma assisted combustion~\cite{starikovskaia2014, popov2016}.

In this paper, we computationally study how streamer properties affect the energy efficiency with which chemically active species are produced.
Simulations are performed of both positive and negative streamers in atmospheric-pressure dry air, using a 2D axisymmetric fluid model.
An advantage of simulations is that they contain information on all species densities and fields, and that discharge parameters such as the voltage waveform can easily be modified.
However, it can be challenging to construct a suitable set of chemical reactions for given conditions and time scales of interest.
We \rev{have here constructed} a set of \rev{263} chemical reactions primarily based on the \rev{reactions from~\cite{kossyi1992, gordiets1995, aleksandrov1999, fresnet2002, tochikubo2002, atkinson2004, florescumitchell2006, gordillo-vazquez2008, popov2011, pancheshnyi2013, ono2020}}, as shown in \ref{sec:chem-reactions}.

\textbf{Past work.} 
Below, we briefly discuss some related computational and experimental work.
Several authors have numerically studied the energy efficiency of species production in pulsed streamer discharges.
This energy efficiency is often reported using $G$-values, in units of atoms or molecules produced per 100\,eV of input energy.
In simulations of positive streamers in air-methane~\cite{bouwman2022}, $G$-values for O and N radicals were found of about 0.7 and 1.5--1.7, respectively.
It was observed that these $G$-values were relatively insensitive to the applied voltage and the streamer length.
In \cite{komuro2015}, the authors simulated the production of N$_2(v$\,=\,1), O($^3$P) and N($^4$S) in primary and secondary streamers in atmospheric-pressure dry air.
When the applied voltage was increased, $G$-values for O($^3$P) increased and $G$-values for N$_2(v$\,=\,1) decreased, since these species have different activation energies.
In \cite{naidis1997a}, $G$-values for the production of oxygen and nitrogen atoms in positive streamers in atmospheric air were computed, in sphere-plane gaps.
It was observed that the calculated $G$-values weakly depended on the applied voltage and discharge conditions, which were found to be about 3--4 and 0.3--0.4 for oxygen and nitrogen atoms, respectively.
An analytical estimate was made for the $G$-values of the production of chemically active species with high activation energies, which was further worked out in~\cite{naidis2012}, as discussed in more detail in section \ref{sec:naidis-estimate}.

We also give a few examples of relevant experimental work on the energy efficiency of NO-removal, which depends on the production of oxygen and nitrogen radicals.
In~\cite{vanveldhuizen1996}, the greatest removal efficiency was found when pulsed positive corona discharges were generated with short high-voltage pulses.
These pulses were shorter than the time required for primary streamers to bridge the discharge gap.
In~\cite{namihira2000}, the authors used pulsed discharges to remove NO in a mixture of N$_2$, O$_2$ and H$_2$O.
They found the removal energy efficiency increased for shorter pulse widths.
The influence of shorter pulse duration on NO-removal energy efficiency was also confirmed by~\cite{matsumoto2010}, in which a high energy efficiency (0.43 mol/kWh) for NO-removal by nanosecond pulsed discharges was found.
Relevant experimental work on the energy efficiency of O$_3$ production can be found in section \ref{sec:comparison-experiment}.

\section{Simulation model}\label{sec:sim-model}

We use a 2D axisymmetric drift-diffusion-reaction type fluid model with the local field approximation to simulate streamers in 80\% $\mathrm{N}_2$ and 20\% $\mathrm{O}_2$, at 300\,K and 1 bar.
Pulsed streamer discharges and their afterglows are simulated up to $t = 500$\,ns using the open-source \verb"Afivo-streamer" code~\cite{teunissen2017}.


\subsection{Model equations}\label{sec:model-equations}

The temporal evolution of the electron density $n_{\mathrm e}$ is given by
\begin{equation}\label{eq:evolution-ne}
    \partial_t n_{\mathrm e} = \nabla \cdot (\mu_{\mathrm e} \boldsymbol{\mathrm E} n_{\mathrm e} + D_{\mathrm e} \nabla n_{\mathrm e}) + S_{\mathrm e} + S_{\mathrm{ph}}\,,
\end{equation}
where $\mu_{\mathrm e}$ is the electron mobility, $\boldsymbol{\mathrm E}$ the electric field, $D_{\mathrm e}$ the electron diffusion coefficient.
Furthermore, $S_{\mathrm{ph}}$ is the source term for non-local photoionization and $S_{\mathrm e}$ is the sum of electron source terms from the reactions listed in table \ref{tab:chem-reactions}.
For photoionization, we use Zheleznyak's model~\cite{zheleznyak1982} and the so-called Helmholtz approximation~\cite{luque2007, bourdon2007}, with the same parameters as in~\cite{guo2022d}.

The temporal evolution of each ion species $n_{\mathrm i}$ ($i = 1, 2, \dots, n$, listed in table \ref{tab:production-all-species}) is given by
\begin{equation}\label{eq:evolution-ni}
    \partial_t n_{\mathrm i} = - \nabla \cdot (\pm \mu_{\mathrm i} \boldsymbol{\mathrm E} n_{\mathrm i}) + S_{\mathrm i}\,,
\end{equation}
where the $\pm$ accounts for the ion charge, $\mu_{\mathrm i}$ the ion mobility, and $S_{\mathrm i}$ is the sum of ion source terms from the reactions listed in table \ref{tab:chem-reactions}.
For simplicity, we use a constant ion mobility $\mu_{\mathrm i}=2.2 \times 10^{-4}$\,m$^2$V$^{-1}$s$^{-1}$~\cite{tochikubo2002} for all ion species, as in~\cite{komuro2012, li2022a}.
For O$_2^+$, the photoionization source term $S_{\mathrm{ph}}$ is included in $S_{\mathrm i}$.

The electric field $\boldsymbol{\mathrm E}$ is calculated as $\boldsymbol{\mathrm E} = - \nabla \phi$. The electric potential $\phi$ is obtained by solving Poisson's equation
\begin{equation}
\label{eq:Poisson-equation}
    \nabla^2 \phi = - \rho/\varepsilon_0\,,
\end{equation}
where $\rho$ is the space charge density and $\varepsilon_0$ is the vacuum permittivity.
Equation (\ref{eq:Poisson-equation}) is solved using the geometric multigrid method included in the Afivo library~\cite{teunissen2018, teunissen2023}.

\subsection{Chemical reactions and input data}\label{sec:reactions-and-input-data}

We have constructed a set of \rev{263} chemical reactions, based on the reactions \rev{from~\cite{kossyi1992, ono2020}} with additional reactions primarily \rev{from~\cite{gordiets1995, aleksandrov1999, fresnet2002, tochikubo2002, atkinson2004, florescumitchell2006, gordillo-vazquez2008, popov2011, pancheshnyi2013}.}
A list of all \rev{56} considered species is given in table \ref{tab:production-all-species}.
The complete reaction list is given in~\ref{sec:chem-reactions}, which also contains a list of the considered excited states of N$_2$ and O$_2$.

The transport coefficients $\mu_{\mathrm e}$ and $D_{\mathrm e}$, and reaction rate coefficients for reactions \rev{R1--R30} in table \ref{tab:chem-reactions} are functions of the reduced electric field $E/N$, where $E$ is the electric field and $N$ is the gas number density.
These coefficients were computed with BOLSIG$+$, a two-term electron Boltzmann equation solver~\cite{hagelaar2005}, using the temporal growth model.
Electron-neutral scattering cross sections for $\mathrm N_2$ and $\mathrm O_2$ were obtained from the Phelps database~\cite{phelps_database, lawton1978, phelps1985}.

\subsection{Computational domain and initial condition}
\label{sec:sim-domain-init-condition}

\begin{figure}
    \centering
    \includegraphics[width=0.48\textwidth]{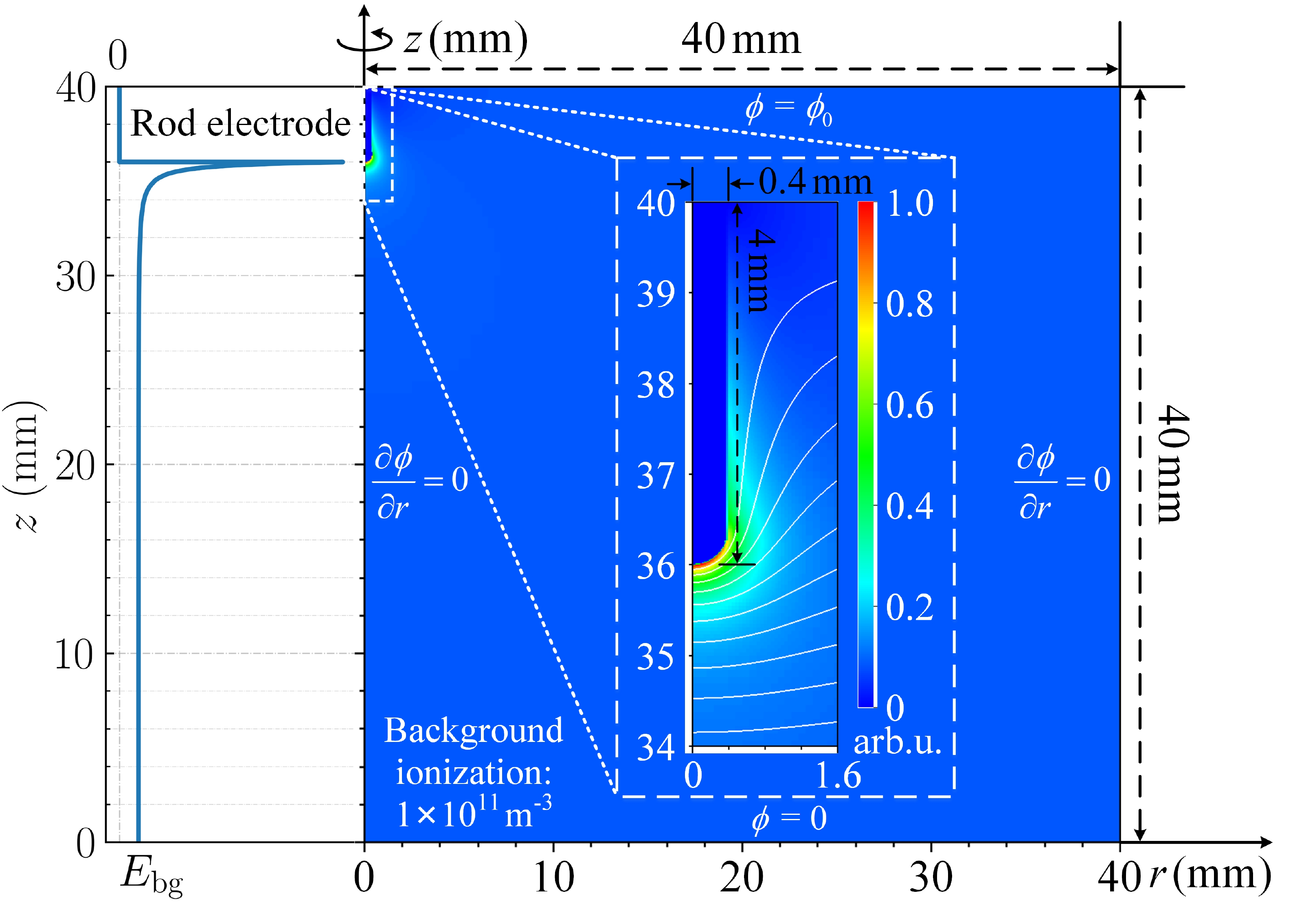}
    \caption{Schematic view of the 2D axisymmetric computational domain, measuring 40\,mm $\times$ 40\,mm.
      Right: the initial electric field without a streamer and the rod electrode geometry for $L_\mathrm{rod}=$ 4\,mm \rev{and $R_\mathrm{rod}=$ 0.4\,mm}.
      Boundary conditions for the electric potential $\phi$ are also indicated.
      Left: the axial electric field.
      $E_\mathrm{bg}$ is the average electric field between two plate electrodes.}
    \label{fig:sim-domain}
\end{figure}

The 2D axisymmetric computational domain used in the simulations is illustrated in figure \ref{fig:sim-domain}. 
The domain measures 40\,mm in the $r$ and $z$ directions, and it contains a plate-plate geometry with a needle protrusion at the upper plate.
To generate non-branching streamers with widely varying properties (such as radius and velocity), we use three different electrodes, \rev{with parameters described in table~\ref{tab:streamer-case}}.
These electrodes are rod-shaped with semi-spherical tips, with lengths $L_\mathrm{rod}$ of 2\,mm, 4\,mm and 6\,mm, and radii \rev{$R_\mathrm{rod}$} of 0.2\,mm, 0.4\,mm and 0.6\,mm, respectively.
Except for a small region near the rod electrode, the axial electric field is approximately uniform and equal to the average electric field between two plate electrodes, which is here defined as the \rev{background electric field $E_\mathrm{bg}$:
\begin{equation}
  \label{eq:bg-field}
  E_\mathrm{bg} = V / d\,,
\end{equation}
where $d = 40$\,mm is the distance between two plate electrodes and $V$ is the applied voltage.}

\begin{table*}
\centering
\caption{\rev{Summary of simulation parameters used for generating 60 positive streamers and 8 negative streamers.}}
\label{tab:streamer-case}
\begin{tabular*}{0.95\textwidth}{c@{\extracolsep{\fill}}ccccc}
  \br
  \multirow{2}{*}{Voltage polarity} & Electrode & Electrode & Steamer & Range of & Number of \\
  & length $L_\mathrm{rod}$ & radius $R_\mathrm{rod}$ & length $L_\mathrm{s}$ & background fields $E_\mathrm{bg}$ & cases \\
  \mr
  \multirow{5}{*}{positive} & 2\,mm & 0.2\,mm & 18\,mm & 11.5--26\,kV/cm & 12 \\
  & 4\,mm & 0.4\,mm & 10\,mm & 10--26\,kV/cm & 12 \\
  & 4\,mm & 0.4\,mm & 18\,mm & 10--26\,kV/cm & 12 \\
  & 4\,mm & 0.4\,mm & 26\,mm & 10--26\,kV/cm & 12 \\
  & 6\,mm & 0.6\,mm & 18\,mm & 9--26\,kV/cm & 12 \\
  \mr
  negative & 6\,mm & 0.6\,mm & 18\,mm & 14--26\,kV/cm & 8 \\
  \br
\end{tabular*}
\end{table*}

For the electric potential, a Dirichlet boundary condition is applied on the upper plate and rod electrode (corresponding to the applied voltage).
The lower plate is grounded, and a homogeneous Neumann boundary condition is applied on the outer axial boundary.
For all plasma species densities, homogeneous Neumann boundary conditions are applied on all domain boundaries, including the rod electrode.
Secondary electron emission from electrodes due to ions and photons is not included.

As an initial condition, homogeneous background ionization with a density of $10^{11}\,\mathrm m^{-3}$ for both electrons and N$_2^+$ is included.
All other ion densities are initially zero.
\rev{The simulations are not sensitive to this background ionization density, because photoionization quickly becomes the dominant source of non-local free electrons after inception, see e.g.~\cite{li2021a, nijdam2011}.}

For computational efficiency, the \texttt{Afivo-streamer} code includes adaptive mesh refinement.
We use the same refinement criteria for the grid spacing $\Delta x$ as~\cite{guo2022d}, which lead to a minimal grid spacing of $\Delta x_\mathrm{min}=$ 1.22\,$\mu$m.

\subsection{Voltage waveform}\label{sec:voltage-waveform}

A single voltage pulse with a rise time of 1\,ns is used, during which it increases linearly.
The applied voltage is then constant until the streamer has reached a desired streamer length~$L_\mathrm{s}$ \rev{(see table~\ref{tab:streamer-case})}, after which the voltage is turned off \rev{linearly} with a 1\,ns fall time.
\rev{The applied voltage $V$ is varied to obtain different background electric fields $E_\mathrm{bg}$, see table~\ref{tab:streamer-case} and equation~(\ref{eq:bg-field}).}

The streamer length $L_\mathrm{s}$ is defined as the distance between the rod electrode tip and the streamer position $z_{\mathrm{head}}$ at which the electric field has a maximum.
We use such a length-dependent voltage waveform so that we can study the effect of $L_\mathrm{s}$ on energy efficiencies (due to more or less Joule heating in the channel, see section~\ref{sec:naidis-estimate}).
Furthermore, it allows us to compare results at the same streamer lengths.
All simulations are performed until $t = 500$\,ns.

\section{Example of streamer dynamics and chemistry}\label{sec:streamer-example}

\subsection{Streamer dynamics}\label{sec:streamer-dynamics}

\begin{figure*}
    \centering
    \includegraphics[width=1\textwidth]{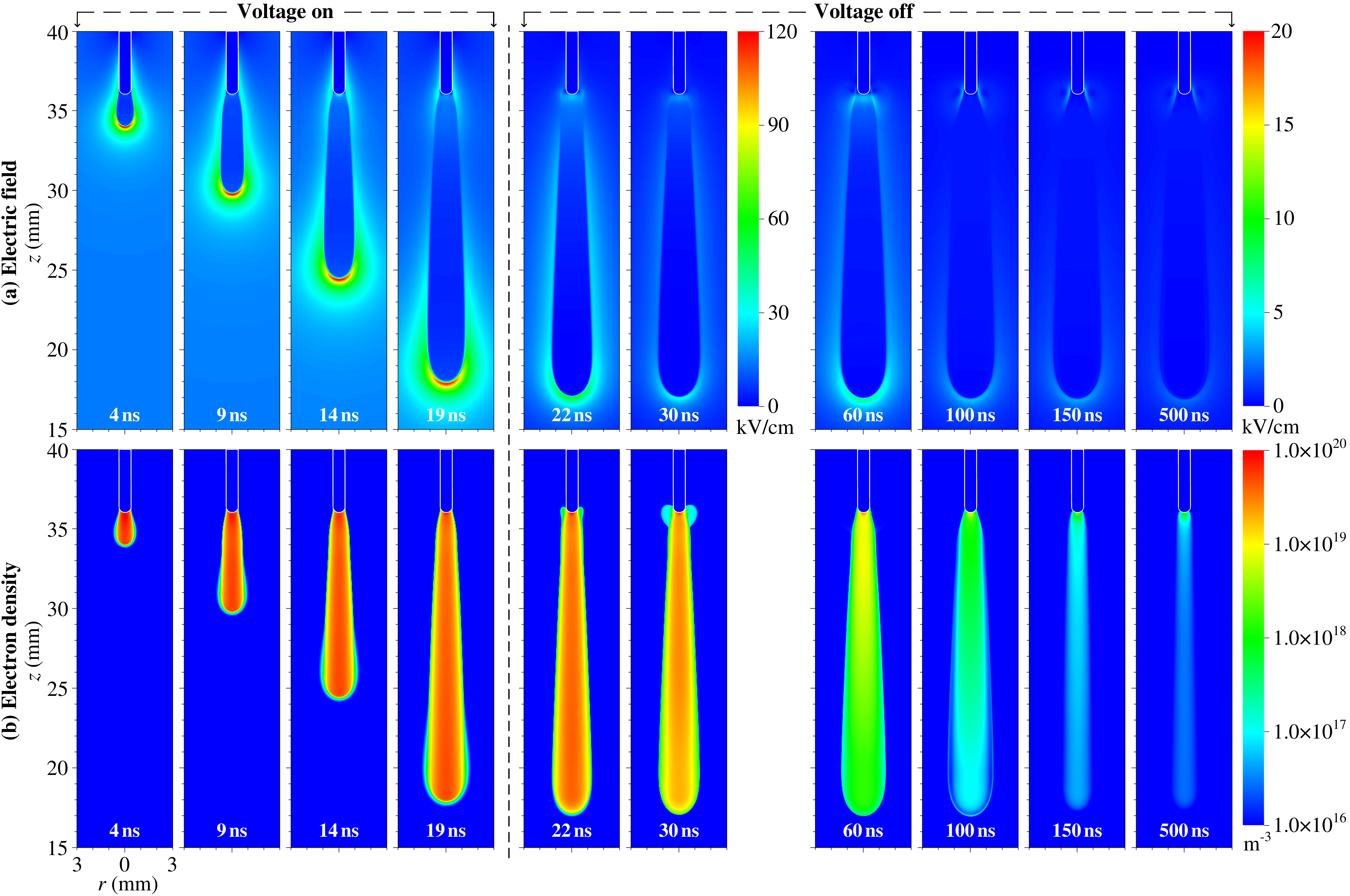}\vspace{0.3em}
    \includegraphics[width=0.95\textwidth]{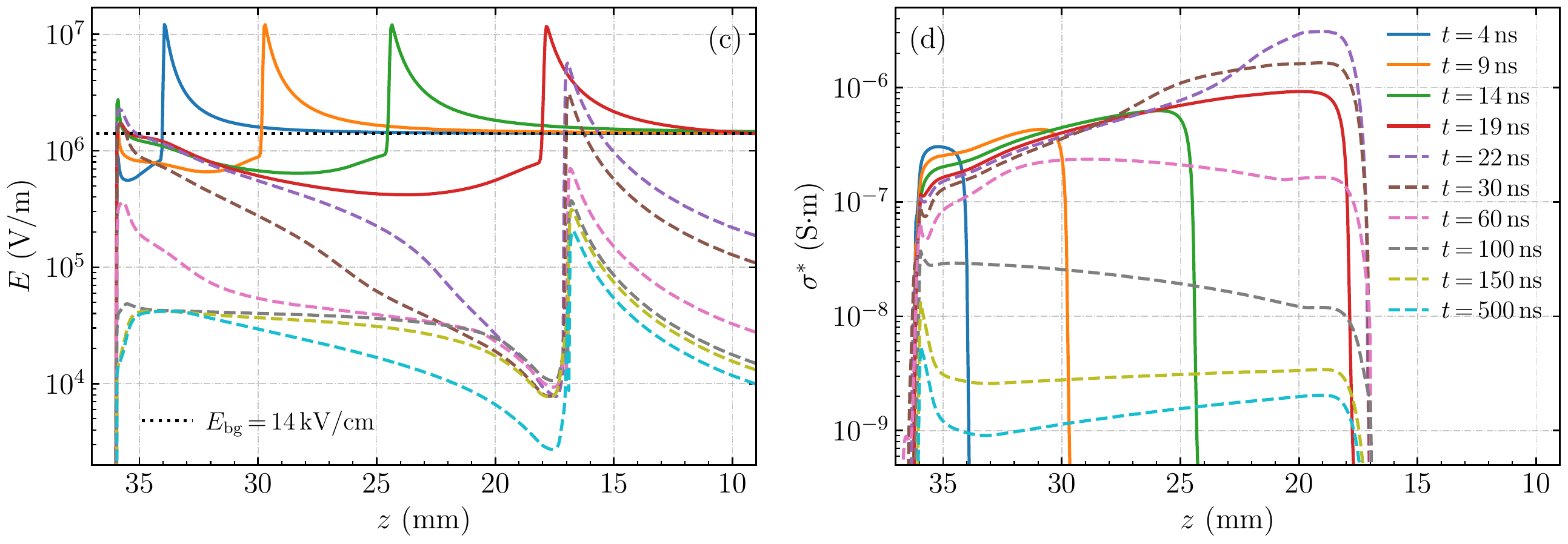}
    \caption{Time evolution of (a) the electric field $E$, (b) the electron density $n_\mathrm{e}$, (c) the on-axis electric field $E$, and (d) the line conductivity $\sigma^*$ for a positive streamer in $E_\mathrm{bg}=14$\,kV/cm with $L_\mathrm{rod}=4$\,mm, \rev{$R_\mathrm{rod}=$ 0.4\,mm} and $L_\mathrm{s}=18$\,mm in air.
    Note that profiles of $E$ after 30\,ns are shown on a different scale; for $n_\mathrm{e}$ the same scale is used at all times.
    Profiles of $n_\mathrm{e}$ are shown on a logarithmic scale.
    All panels in (a) and (b) are zoomed in into the region where 0 $\leqslant r \leqslant$ 3\,mm and 15 $\leqslant z \leqslant$ 40\,mm.
    The solid and dashed lines in panels (c) and (d) correspond to the streamer before and after the voltage is turned off, respectively.}
    \label{fig:E-ne-evolution-axial-profiles}
\end{figure*}

We first present an example of a positive streamer in dry air.
A background field $E_\mathrm{bg}$ of 14\,kV/cm (half of the breakdown field of 28\,kV/cm in air) was used, and the voltage was turned off after \rev{19.8\,ns}.
Furthermore, a rod electrode \rev{with length $L_\mathrm{rod}=4$\,mm and radius $R_\mathrm{rod}=0.4$\,mm} and a desired streamer length $L_\mathrm{s}=18$\,mm were used.
Figure \ref{fig:E-ne-evolution-axial-profiles} shows the time evolution of the electric field and electron density profiles for this example, together with its axial profiles.

The streamer initiates from the rod electrode tip, after which its velocity and radius increase almost linearly with time, from about $0.7\times10^6$\,m/s to $1.4\times10^6$\,m/s and from about 0.5\,mm to 0.9\,mm, respectively.
The streamer radius is here defined as the electrodynamic radius at which the radial component of the electric field has a maximum.

The line conductivity $\sigma^*$ at the streamer head also increases as the streamer grows, due to its increasing radius, from about $3\times10^{-7}$\,S$\cdot$m to $3\times10^{-6}$\,S$\cdot$m.
Here $\sigma^*$ is computed as
\begin{equation}
\label{eq:line-conductivity}
  \sigma^*(z) = 2 \pi e \int_0^{10\,\mathrm{mm}} r n_\mathrm{e} \mu_\mathrm{e} \, \mathrm{d} r\,,
\end{equation}
where $e$ is the elementary charge.
In contrast, the electron density at the streamer head decreases slightly from about $6\times10^{19}$\,m$^{-3}$ to $5\times10^{19}$\,m$^{-3}$, \rev{and the lowest internal on-axis electric field inside the streamer channel decreases from about 6\,kV/cm to 4\,kV/cm. 
However, the maximal electric field at the streamer head of about 120\,kV/cm stays approximately constant during the voltage pulse.}

After \rev{19.8\,ns} the voltage is turned off and the streamer stops, but the simulation continues up to 500\,ns.
The streamer channel gradually loses its conductivity, with an \rev{about three} orders of magnitude decrease in the electron density and line conductivity.
The channel radius increases slightly from about 0.9\,mm to 1.2\,mm due to ion motion.

At 500\,ns, the maximal electric field is about \rev{2\,kV/cm}, and the lowest internal field is about \rev{0.03\,kV/cm}.
The total deposited energy $Q_\mathrm{total}$ at 500\,ns is about \rev{86\,$\mu$J}, which is here computed as
\begin{equation}
\label{eq:input-energy}
  Q_\mathrm{total} = \int_{0}^{T_\mathrm{total}}\int_{\Omega} j \cdot E \,\mathrm{d} \Omega \mathrm{d} t\,,
\end{equation}
where $j = e n_\mathrm{e} \mu_\mathrm{e} E$ is the electron conduction current density, $\Omega$ the computational domain, and $T_\mathrm{total}=500$\,ns.

\subsection{Plasma chemistry}\label{sec:plasma-chemistry}

We now look into the plasma chemistry of the streamer case mentioned above.
We consider \rev{56 chemically active species and 263} chemical reactions, as shown in table \ref{tab:chem-reactions}.
A list of \emph{gross} and \emph{net} productions of all \rev{56} species at $t=500$\,ns is given in table \ref{tab:production-all-species}, in units of the number of molecules/atoms produced in some particular state.
Here gross production means the total time and space integrated production of a specific species, without taking into account loss processes.
For net production, loss processes are taken into account.

\begin{table}
\renewcommand{\baselinestretch}{1.02}
\footnotesize
\centering
\caption{\rev{The gross and net productions of all 56 species for the streamer corresponding to figure \ref{fig:E-ne-evolution-axial-profiles} at $t=500$\,ns. Production below $10^3$ has been replaced by $\sim 0$.}}
\label{tab:production-all-species}
\lineup
\begin{tabular*}{0.49\textwidth}{l@{\extracolsep{\fill}}lll}
\br
Species & Gross production & Net production \\
\mr
N$_2(J)$ & $2.60\times10^{14}$ & $2.60\times10^{14}$ \\
N$_2(v_1)$ & $3.50\times10^{14}$ & $3.50\times10^{14}$ \\
N$_2(v_2)$ & $1.39\times10^{14}$ & $1.39\times10^{14}$ \\
N$_2(v_3)$ & $7.91\times10^{13}$ & $7.91\times10^{13}$ \\
N$_2(v_4)$ & $2.86\times10^{13}$ & $2.86\times10^{13}$ \\
N$_2(v_5)$ & $1.78\times10^{13}$ & $1.78\times10^{13}$ \\
N$_2(v_6)$ & $7.92\times10^{12}$ & $7.92\times10^{12}$ \\
N$_2(v_7)$ & $2.83\times10^{12}$ & $2.83\times10^{12}$ \\
N$_2(v_8)$ & $9.96\times10^{11}$ & $9.96\times10^{11}$ \\
\rev{N$_2$(A)} & $2.96\times10^{12}$ & $3.03\times10^{8}$ \\
N$_2$(B) & $5.44\times10^{12}$ & $\sim 0$ \\ 
N$_2$(a) & $2.66\times10^{12}$ & $\sim 0$ \\ 
N$_2$(C) & $2.48\times10^{12}$ & $\sim 0$ \\ 
N$_2$(E) & $1.69\times10^{11}$ & $\sim 0$ \\ 
\mr
O$_2(J)$ & $3.03\times10^{12}$ & $3.03\times10^{12}$ \\
O$_2(v_1)$ & $3.41\times10^{13}$ & $3.41\times10^{13}$ \\
O$_2(v_2)$ & $9.58\times10^{12}$ & $9.58\times10^{12}$ \\
O$_2(v_3)$ & $2.61\times10^{12}$ & $2.61\times10^{12}$ \\
O$_2(v_4)$ & $8.04\times10^{11}$ & $8.04\times10^{11}$ \\
O$_2$(a) & $2.56\times10^{12}$ & $2.54\times10^{12}$ \\
O$_2$(b) & $3.15\times10^{12}$ & $3.09\times10^{12}$ \\
O$_2$(A) & $1.09\times10^{12}$ & $1.71\times10^{11}$ \\
\mr
N($^4$S) & $2.38\times10^{12}$ & $2.38\times10^{12}$ \\
N($^2$D) & $2.38\times10^{12}$ & $1.66\times10^{5}$ \\
N($^2$P) & $4.57\times10^{8}$ & $2.91\times10^{6}$ \\
O($^3$P) & $2.95\times10^{13}$ & $2.83\times10^{13}$ \\
O($^1$D) & $5.48\times10^{12}$ & $7.08\times10^{8}$ \\
O($^1$S) & $2.16\times10^{12}$ & $1.16\times10^{12}$ \\
O$_3$ & $1.10\times10^{12}$ & $1.10\times10^{12}$ \\
NO & $2.37\times10^{12}$ & $2.36\times10^{12}$ \\
NO$_2$ & $6.94\times10^{9}$ & $6.89\times10^{9}$ \\
NO$_3$ & $5.30\times10^{6}$ & $5.18\times10^{6}$ \\
N$_2$O & $6.09\times10^{10}$ & $6.09\times10^{10}$ \\
N$_2$O$_3$ & $\sim 0$ & $\sim 0$ \\
N$_2$O$_4$ & $\sim 0$ & $\sim 0$ \\
N$_2$O$_5$ & $6.53\times10^{3}$ & $6.53\times10^{3}$ \\
\mr
$e$ & $1.23\times10^{12}$ & $7.17\times10^{8}$ \\
O$^-$ & $8.91\times10^{10}$ & $5.60\times10^{8}$ \\
O$_2^-$ & $8.02\times10^{12}$ & $2.66\times10^{11}$ \\
O$_3^-$ & $9.70\times10^{10}$ & $5.73\times10^{10}$ \\
O$_4^-$ & $7.36\times10^{12}$ & $1.46\times10^{11}$ \\
NO$^-$ & $1.60\times10^{7}$ & $1.60\times10^{7}$ \\
NO$_2^-$ & $2.50\times10^{7}$ & $2.50\times10^{7}$ \\
NO$_3^-$ & $1.22\times10^{9}$ & $1.22\times10^{9}$ \\
N$_2$O$^-$ & $5.79\times10^{4}$ & $5.79\times10^{4}$ \\
N$^+$ & $4.34\times10^{5}$ & $4.34\times10^{5}$ \\
N$_2^+$ & $8.96\times10^{11}$ & $2.23\times10^{6}$ \\
N$_3^+$ & $1.49\times10^{6}$ & $1.49\times10^{6}$ \\
N$_4^+$ & $1.27\times10^{12}$ & $\sim 0$ \\ 
O$^+$ & $4.70\times10^{7}$ & $4.70\times10^{7}$ \\
O$_2^+$ & $2.27\times10^{12}$ & $1.11\times10^{9}$ \\
O$_4^+$ & $1.44\times10^{12}$ & $4.69\times10^{11}$ \\
NO$^+$ & $1.62\times10^{9}$ & $1.62\times10^{9}$ \\
NO$_2^+$ & $7.72\times10^{6}$ & $7.72\times10^{6}$ \\
N$_2$O$^+$ & $\sim 0$ & $\sim 0$ \\
N$_2$O$_2^+$ & $1.20\times10^{12}$ & $2.91\times10^{7}$ \\
\br
\end{tabular*}
\end{table}

\begin{figure*}
  \centering
  \includegraphics[width=1\textwidth]{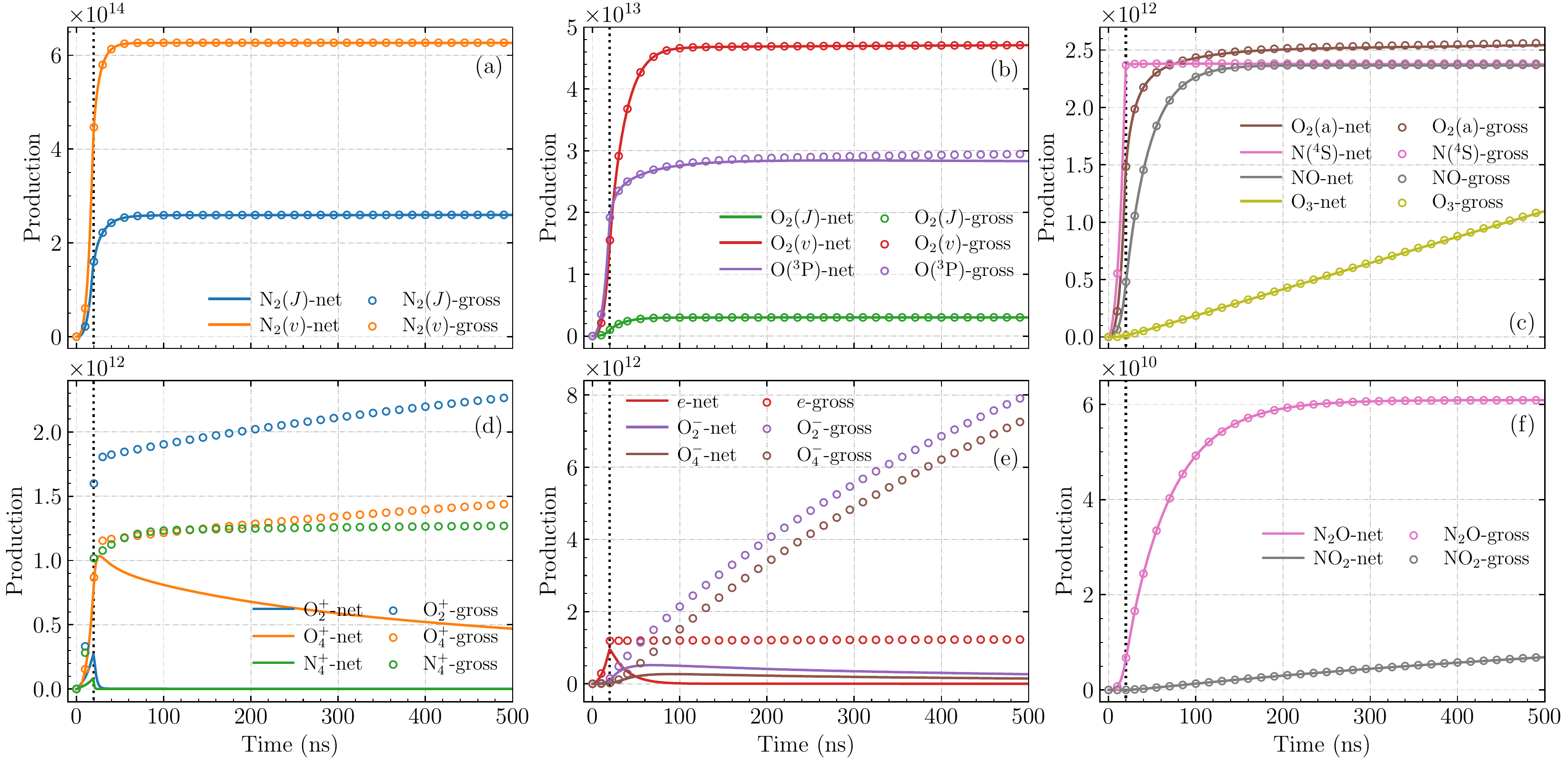}
  \caption{Time evolution of the gross and net productions of 17 species for the streamer corresponding to figure \ref{fig:E-ne-evolution-axial-profiles}.
  The vertical dotted black lines correspond to the moment when the voltage is turned off, namely \rev{$t=19.8$\,ns}.}
  \label{fig:production-vs-time}
\end{figure*}

The time evolution of the gross and net production of 17 species is shown in figure \ref{fig:production-vs-time}, namely $e$ (electrons), N$_2(J)$, N$_2(v)$, O$_2(J)$, O$_2(v)$, O$_2$(a), N($^4$S), O($^3$P), O$_2^+$, O$_4^+$, N$_4^+$, O$_2^-$, O$_4^-$, O$_3$, NO, NO$_2$ and N$_2$O.
Note that the total vibrationally excited state N$_2(v)$ is the sum of each vibrationally excited state from N$_2(v_1)$ to N$_2(v_8)$, and the total vibrationally excited state O$_2(v)$ is the sum of each vibrationally excited state from O$_2(v_1)$ to O$_2(v_4)$.

Figure \ref{fig:production-vs-time} shows that the gross and net production are the same for N$_2(J)$, N$_2(v)$, O$_2(J)$ and O$_2(v)$, because their loss reactions were not included, \rev{as discussed in \ref{sec:reaction-set-comment}.}
For O$_2$(a), N($^4$S), O($^3$P), O$_3$, NO, NO$_2$ and N$_2$O, gross and net production are almost equal due to relatively slow loss processes.
\rev{The gross and net production of O($^3$P) differ because O($^3$P) is primarily converted to O$_3$ by the reaction $\rm O(^3P) + O_2 + M \to O_3 + M$.
}
For charged species, net production is much lower than gross production since these species are rapidly converted to other species by attachment, detachment, ion conversion and recombination.

For most of the 17 species there is essentially no more production after 500\,ns, with the exceptions being O$_3$, NO$_2$, \rev{O$_2^+$, O$_4^+$, N$_4^+$,} O$_2^-$ and O$_4^-$, which are produced relatively slowly.
\rev{Note that there is continued gross production of both O$_2^-$ and O$_4^-$, but no net production, due to the conversion between these species by reactions R69 and R81.}

\section{Parameter study of species production and energy efficiency}\label{sec:energy-efficiency}

\begin{figure*}
  \centering
  \includegraphics[width=1\textwidth]{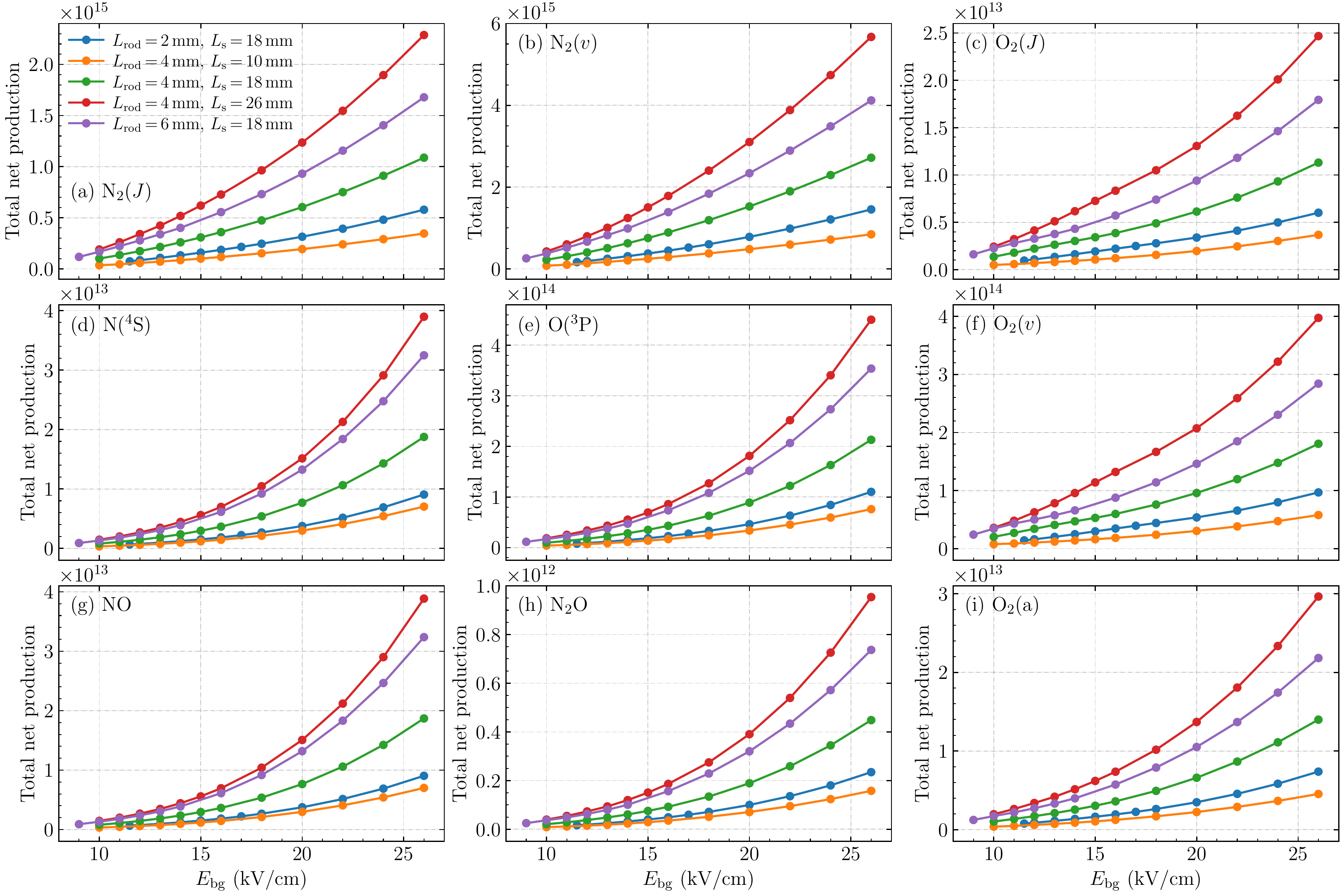}
  \caption{The total net production at 500\,ns of nine species versus the background field $E_\mathrm{bg}$.
  Results are shown for 60 positive streamers in different background fields with different electrode \rev{geometries ($L_\mathrm{rod}=2, 4, 6$\,mm, $R_\mathrm{rod}=0.2, 0.4, 0.6$\,mm)} and different streamer lengths ($L_\mathrm{s}=10, 18, 26$\,mm), \rev{see table~\ref{tab:streamer-case}}.}
  \label{fig:nine-net-production-vs-Ebg}
\end{figure*}
\begin{figure*}
  \centering
  \includegraphics[width=1\textwidth]{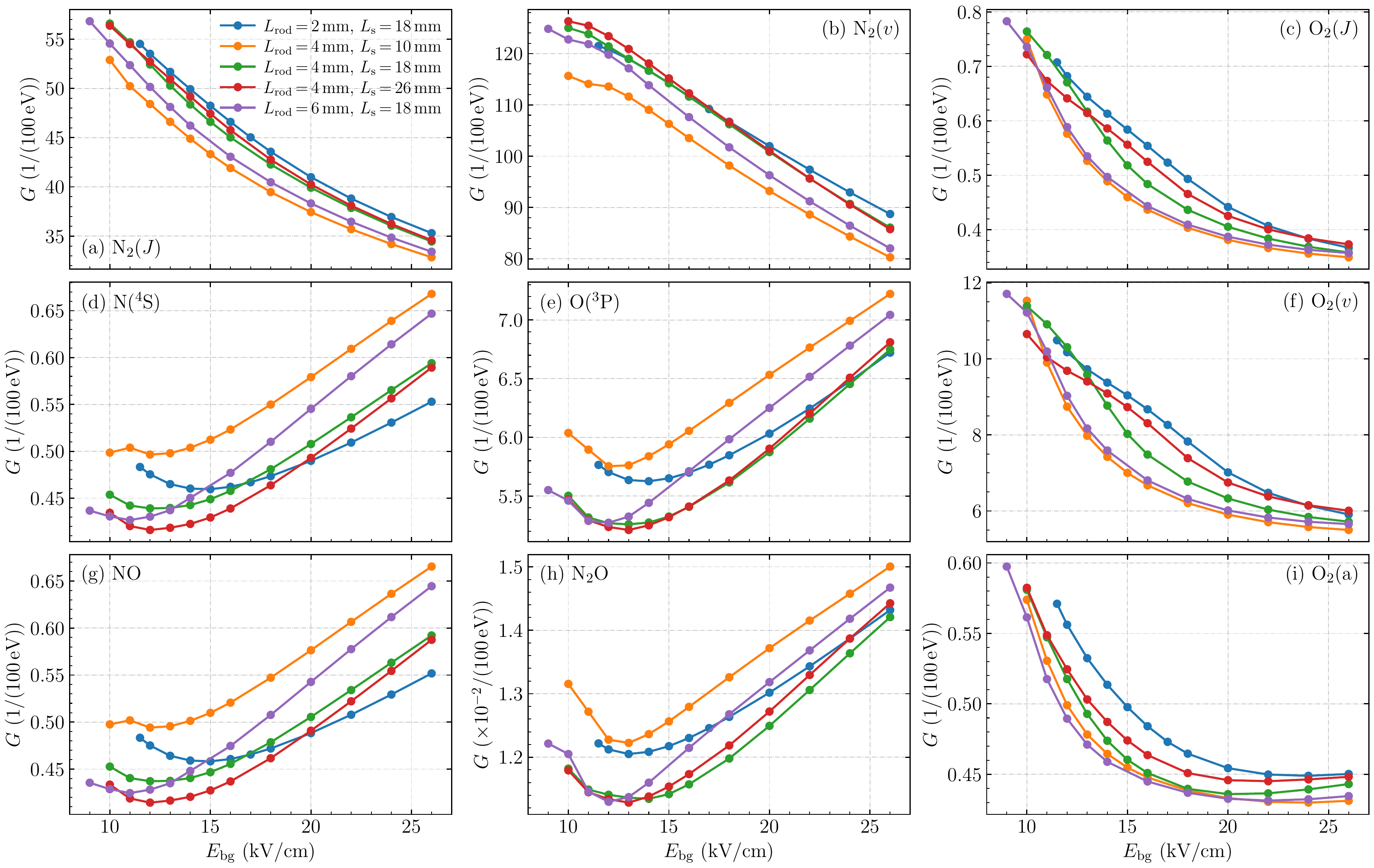}
  \caption{$G$-values for the production of nine species versus the background field $E_\mathrm{bg}$ for the 60 streamers corresponding to figure \ref{fig:nine-net-production-vs-Ebg}.}
  \label{fig:nine-G-net-vs-Ebg}
\end{figure*}

In this section, we simulate 60 positive streamers by \rev{varying the rod electrode length, rod electrode radius, desired streamer length and the background field as described in table~\ref{tab:streamer-case}.}
In figures \ref{fig:nine-net-production-vs-Ebg} and \ref{fig:nine-G-net-vs-Ebg}, we compare the total net production and energy efficiencies for nine neutral species: N$_2(J)$, N$_2(v)$, O$_2(J)$, O$_2(v)$, O$_2$(a), N($^4$S), O($^3$P), NO and N$_2$O.
To compare energy efficiencies, we use so-called $G$-values, which give the net number of atoms or molecules produced per 100\,eV of deposited energy, at $t=500$\,ns.
We find the highest $G$-values for N$_2(v)$, about $10^2$, and the lowest for N$_2$O, about $10^{-2}$.

Figure \ref{fig:nine-net-production-vs-Ebg} shows that the total net production of the nine species increases with the background field and with the streamer length, as could be expected.
A longer rod electrode leads to a wider and faster streamer and therefore also to more production.

Figure \ref{fig:nine-G-net-vs-Ebg} shows that the $G$-values for N$_2(v)$, O$_2$(a), N($^4$S), O($^3$P), NO and N$_2$O vary by at most 30\% to \rev{60\%}.
Larger variation of about 70\% to \rev{120\%} can be observed for N$_2(J)$, O$_2(J)$ and O$_2(v)$.
The dependence of the $G$-value on the background field for nine species can be grouped into three categories.
For N($^4$S), O($^3$P), NO and N$_2$O, $G$-values first slightly decrease and then increase with the background field, whereas for N$_2(J)$, N$_2(v)$, O$_2(J)$ and O$_2(v)$ $G$-values monotonically decrease with the background field.
For O$_2$(a), $G$-values first decrease and then slightly increase with the background field.
These different dependencies can be explained by considering the activation energies for the reactions producing these species.

\begin{figure}
  \centering
  \includegraphics[width=0.48\textwidth]{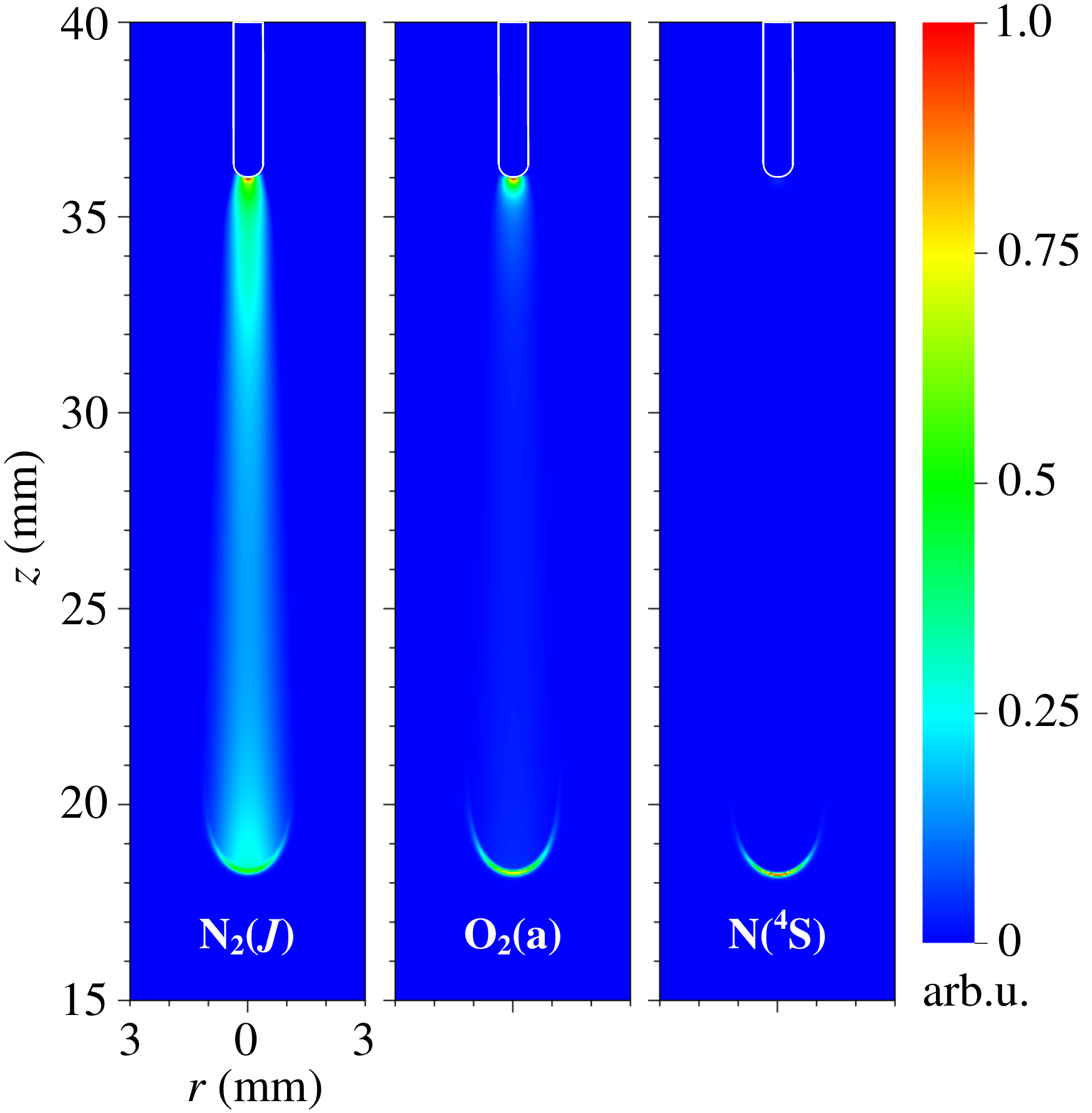}
  \caption{The instantaneous production of N$_2(J)$, O$_2$(a) and N($^4$S) at \rev{$t=18.8$\,ns for the streamer corresponding to figure \ref{fig:E-ne-evolution-axial-profiles}.}
    N$_2(J)$ is mostly produced in the channel, N($^4$S) mostly in the head and O$_2$(a) is produced in both.
    Profiles are shown using arbitrary units.}
  \label{fig:source-term}
\end{figure}

The key reactions producing N($^4$S), O($^3$P), NO and N$_2$O have high activation energies ranging from 6\,eV to 13\,eV, so they are primarily produced in the high electric field near the streamer head, as illustrated in figure \ref{fig:source-term} for N($^4$S).
Specifically, N($^4$S) and NO are mostly produced by an electron dissociation reaction, namely $\rm e + N_2 \to e + N(^4S) + N(^2D)$, and
the N$_2$O production depends on electronic excitation reactions producing \rev{N$_2$(A)} and N$_2$(B).
O($^3$P) production is determined by electron dissociation reactions of nitrogen and oxygen molecules, as well as electronic excitation reactions of nitrogen, e.g., N$_2$(B).

\begin{figure}
  \centering
  \includegraphics[width=0.48\textwidth]{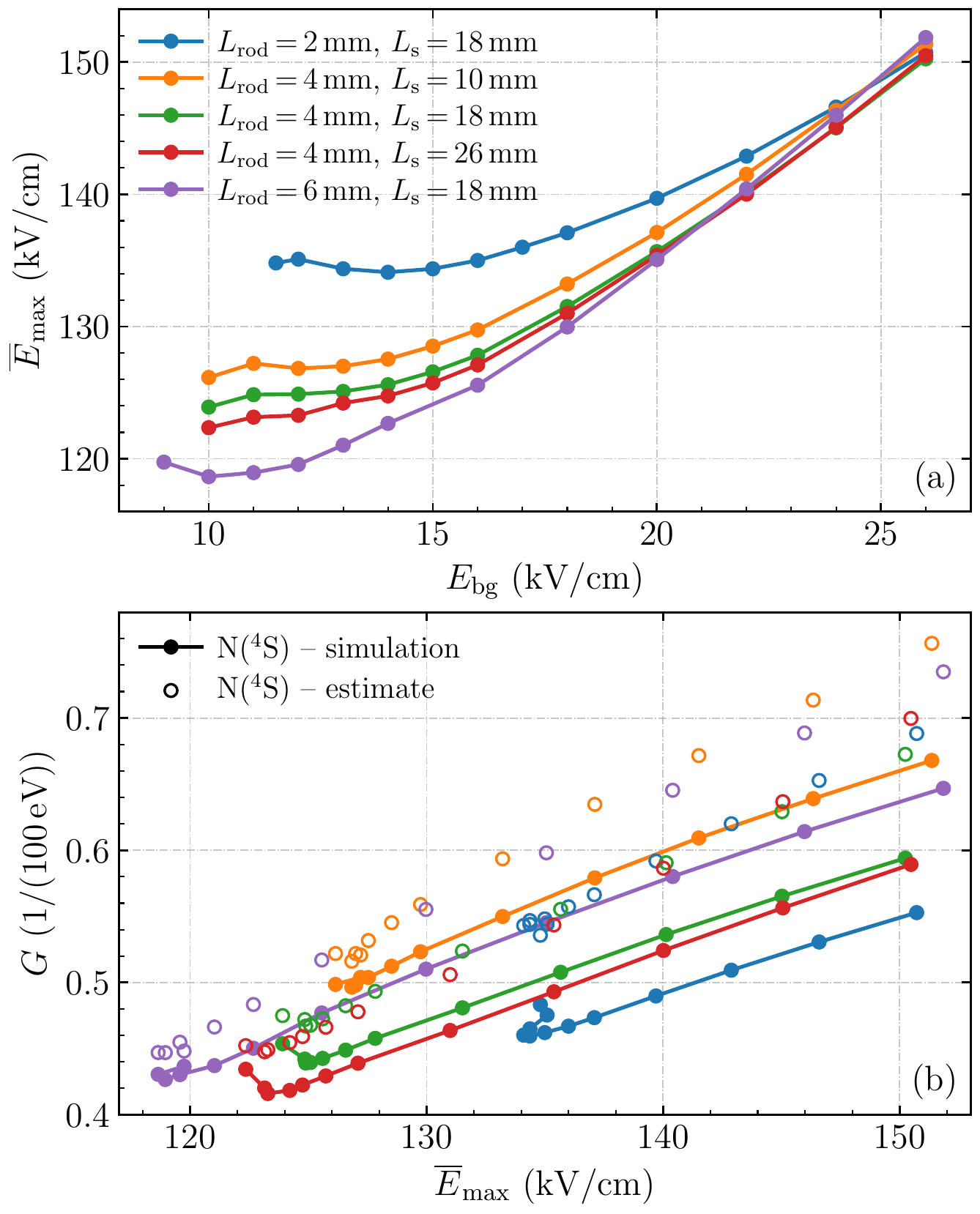}
  \caption{(a) The average maximal electric field $\overline{E}_\mathrm{max}$ versus the background field $E_\mathrm{bg}$ for the 60 streamers corresponding to figure \ref{fig:nine-net-production-vs-Ebg}.
  (b) The simulated and estimated $G$-values for N($^4$S) versus $\overline{E}_\mathrm{max}$ corresponding to panel (a).
  The symbols correspond to the 60 simulated cases, by taking an average over the time when the voltage is at its peak for $\overline{E}_\mathrm{max}$, by taking simulated $G$-values at $t=500$\,ns and by taking estimated $G$-values from equation (\ref{eq:G-value-estimate}) corrected with the factor $Q_\mathrm{head}/Q_\mathrm{total}$ from figure~\ref{fig:head-energy-ratio}(d).}
  \label{fig:N4S-G-comparison}
\end{figure}

Figure \ref{fig:N4S-G-comparison}(a) shows the relation between the average maximal electric field $\overline{E}_\mathrm{max}$ and the background field $E_\mathrm{bg}$ for the 60 streamer cases.
Here $\overline{E}_\mathrm{max}$ is an average over the time when the voltage is at its peak.
\rev{Note that $\overline{E}_\mathrm{max}$ depends non-monotonically on the background field: it increases for both the highest and lowest considered background fields, as was also observed in~\cite{francisco2021e}.
This explains the similarly non-monotonic dependence of the $G$-values for N($^4$S), O($^3$P), NO and N$_2$O shown in figure~\ref{fig:nine-G-net-vs-Ebg}, since there is an approximately linear relation between $\overline{E}_\mathrm{max}$ and these $G$-values, as illustrated in figure~\ref{fig:N4S-G-comparison}(b).}
\rev{We remark that it was difficult to avoid streamer branching in simulations in low background fields, so we could not fully explore the increase in $G$-values for low background fields.}

In contrast, the electron-molecule reactions producing N$_2(J)$, N$_2(v)$, O$_2(J)$ and O$_2(v)$ have lower activation energies, below about 2\,eV.
These reactions therefore primarily take place inside the streamer channel, where the electron density is high, as illustrated in figure \ref{fig:source-term} for N$_2(J)$.
Specifically, N$_2(J)$ and O$_2(J)$ are produced by rotational excitations of nitrogen and oxygen
molecules, and N$_2(v)$ and O$_2(v)$ by vibrational excitations.
Figure \ref{fig:nine-net-production-vs-Ebg} shows that the production of N$_2(J)$, N$_2(v)$, O$_2(J)$ and O$_2(v)$ increases for higher background fields, but that the production of N($^4$S), O($^3$P), NO and N$_2$O increases even more rapidly, so that $G$-values for N$_2(J)$, N$_2(v)$, O$_2(J)$ and O$_2(v)$ decrease.

The production of O$_2$(a) falls somewhat between these two regimes, because it can be produced by reactions with different activation energies.
There is a direct electronic excitation reaction generating O$_2$(a) with an activation energy of 0.977\,eV, but O$_2$(a) is also produced indirectly from \rev{N$_2$(A)} and N$_2$(B), which correspond to higher activation energies, ranging from 6.17\,eV to 8.16\,eV.
Due to these different mechanisms, O$_2$(a) is both produced inside the streamer channel and in the high electric field near the streamer head, as shown in figure \ref{fig:source-term}.

\section{Other results and discussion}\label{sec:discussion}

\subsection{Comparison with Naidis' estimate}\label{sec:naidis-estimate}

In~\cite{naidis2012}, Naidis proposed that ``the maximum $G$-values are governed by only one parameter -- the reduced electric field in the streamer head $Y_\mathrm{h}$, and that their dependence on $Y_\mathrm{h}$ in the typical range of propagating streamers is not strong, when both generation of active species and energy input occur mainly in the streamer head region.''
The paper contains an analytic approximation for $G$-values, that we for clarity rewrite as:
\begin{equation}
\label{eq:G-value-estimate}
  G_j = \frac{2 n_j}{e E_\mathrm{max}^2} \int_{E_\mathrm{ch}}^{E_\mathrm{max}} \frac{K_j}{\mu_\mathrm{e} E} \, \mathrm{d} E \,,
\end{equation}
where $E_\mathrm{max}$ is the maximal electric field at the streamer head, $E_\mathrm{ch}$ the electric field in the streamer channel, $n_j$ the number density of molecules producing species $j$, $K_j$ the field-dependent rate coefficient of electron-molecule reaction producing species $j$, and $E$ is the electric field.
The integral is not sensitive to the particular value used for $E_\mathrm{ch}$, since it is meant to be used for species mainly produced in high electric fields, so we for simplicity use $E_\mathrm{ch} = 0$ below. 
Several approximations are made to derive equation~(\ref{eq:G-value-estimate}): the streamer is assumed to be uniformly translating, the streamer head is locally assumed to be flat, so that one-dimensional integration can be performed, and Joule heating in the streamer channel is not taken into account.

\begin{figure}
  \centering
  \includegraphics[width=0.48\textwidth]{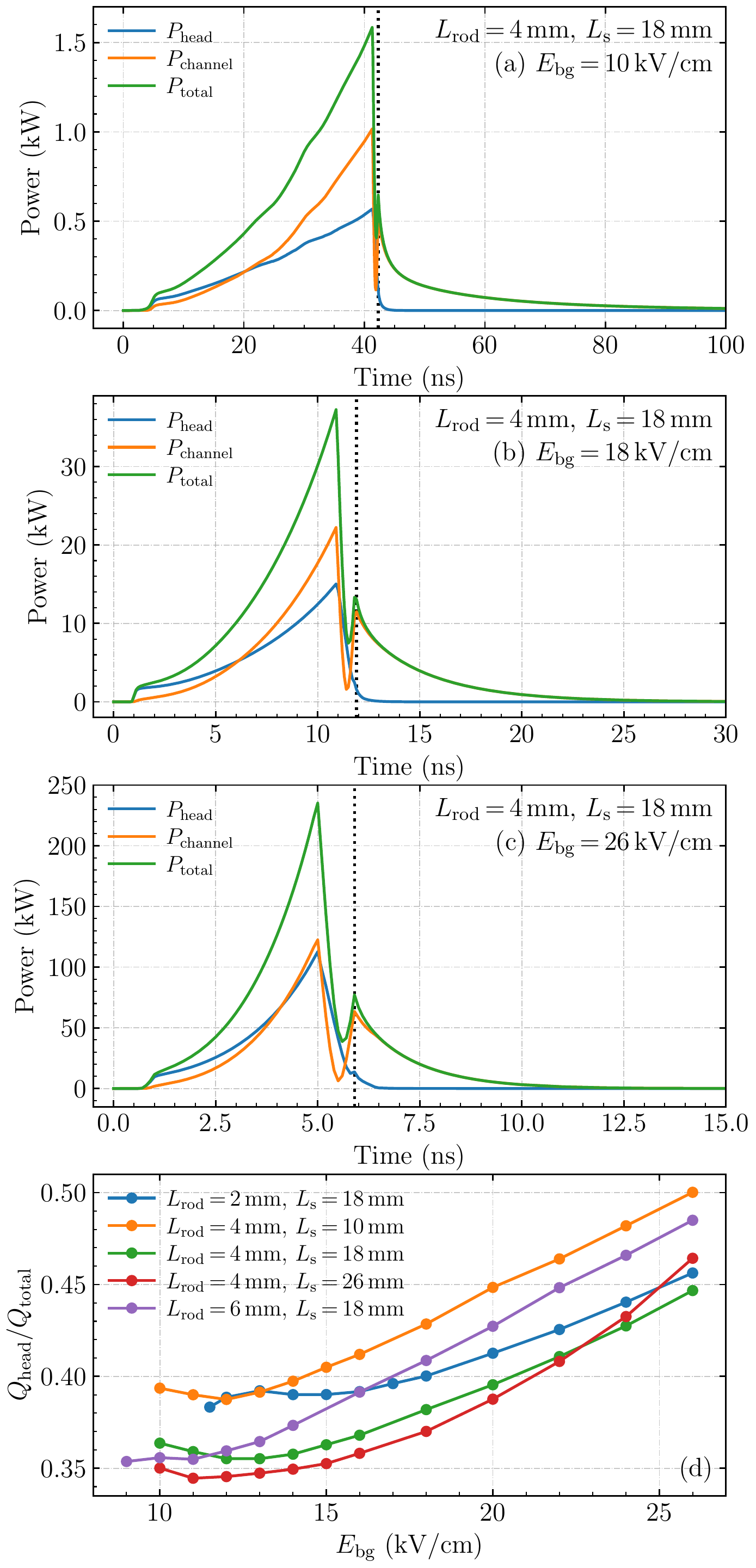}
  \caption{Time evolution of total deposited power $P_\mathrm{total}$, power in the streamer head region $P_\mathrm{head}$ and in the streamer channel $P_\mathrm{channel}$ for three positive streamers in $E_\mathrm{bg}$ of (a) 10\,kV/cm, (b) 18\,kV/cm and (c) 26\,kV/cm with $L_\mathrm{rod}=4$\,mm, \rev{$R_\mathrm{rod}=$ 0.4\,mm} and $L_\mathrm{s}=18$\,mm.
  The vertical dotted black lines correspond to the moment when the applied voltage has dropped to zero (after a 1\,ns fall time).
  The streamer head region is defined as the region where $E \geqslant 28$\,kV/cm.
  (d) The fraction of energy deposited in the streamer head region $Q_\mathrm{head}/Q_\mathrm{total}$ versus the background field $E_\mathrm{bg}$ for the 60 streamers corresponding to figure \ref{fig:nine-net-production-vs-Ebg}.}
  \label{fig:head-energy-ratio}
\end{figure}

As discussed in~\cite{naidis2012}, it can be important to take Joule heating in the streamer channel into account.
We obtain this correction by measuring the deposited power in the streamer channel and in the streamer head region, as shown in figure~\ref{fig:head-energy-ratio}.
The streamer head region was defined as the region where $E \geqslant 28$\,kV/cm.
There is still some Joule heating in the streamer channel after the voltage is turned off, which corresponds to currents that screen the electric field generated by the remaining space charge.
The fraction of power deposited in the channel increases with time.
However, even for short primary streamers, on time scales well below 10\,ns, Joule heating in the streamer channel can still be important.

A correction factor $Q_\mathrm{head}/Q_\mathrm{total}$, i.e., the fraction of total energy deposited in the streamer head region, is computed by integrating the deposited power over time.
As shown in figure \ref{fig:head-energy-ratio}(d), this factor lies between 0.35 and 0.5 for the 60 positive cases considered in this paper.
Note that $Q_\mathrm{head}/Q_\mathrm{total}$ first decreases and then increases with the background field, similar to $\overline{E}_\mathrm{max}$ shown in figure \ref{fig:N4S-G-comparison}(a).

In figure \ref{fig:N4S-G-comparison}(b), we compare equation~(\ref{eq:G-value-estimate}) corrected with the factors $Q_\mathrm{head}/Q_\mathrm{total}$ from figure~\ref{fig:head-energy-ratio}(d) against simulation results, for the production of N($^4$S).
For this comparison, we use $E_\mathrm{max} = \overline{E}_\mathrm{max}$ and $E_\mathrm{ch} = 0$ in equation~(\ref{eq:G-value-estimate}).
The agreement is then surprisingly good, with deviations of up to about 25\%. 
We remark that the variation of equation (\ref{eq:G-value-estimate}) is about 20\% in the considered maximal field range, which is smaller than the variation in the correction factor from figure~\ref{fig:head-energy-ratio}(d), which is about \rev{40\%}. 


\subsection{Comparison with negative streamers}\label{sec:comparison-negative}

We now compare the species production and $G$-values for N($^4$S), O($^3$P) and NO between positive and negative streamers in different background fields, \rev{as described in table~\ref{tab:streamer-case}}.
A rod electrode \rev{with length $L_\mathrm{rod}=6$\,mm and radius $R_\mathrm{rod}=0.6$\,mm} and a desired streamer length $L_\mathrm{s}=18$\,mm were used.
Positive and negative streamers were obtained by changing the applied voltage polarity, while keeping all other simulation conditions the same.

\begin{figure}
  \centering
  \includegraphics[width=0.48\textwidth]{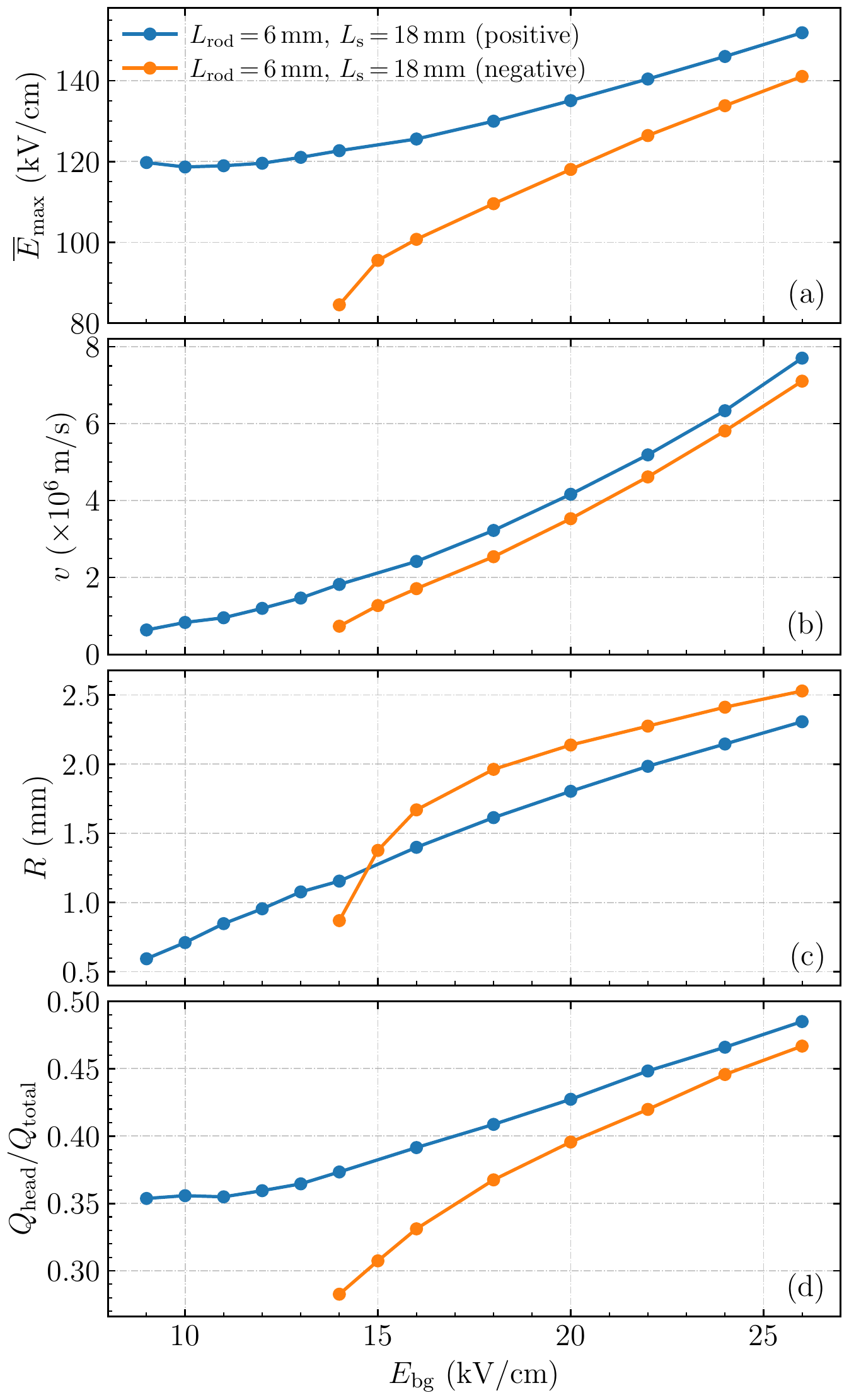}
  \includegraphics[width=0.48\textwidth]{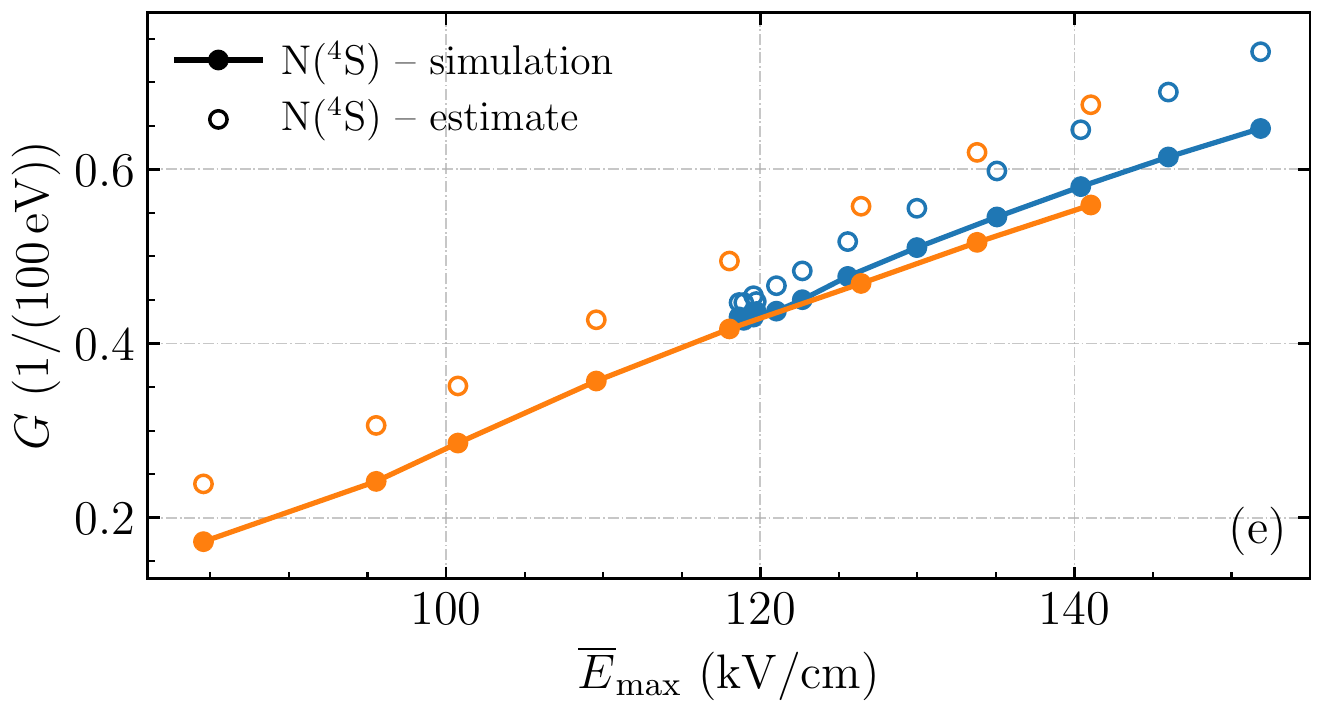}
  \caption{Comparison of positive and negative streamers, \rev{with parameters described in table~\ref{tab:streamer-case}}.
  (a) Time-averaged maximal electric field $\overline{E}_\mathrm{max}$, (b) velocity $v$, (c) radius $R$ and (d) fraction of energy deposited in the streamer head region $Q_\mathrm{head}/Q_\mathrm{total}$, all versus the background field $E_\mathrm{bg}$.
  The velocity and radius are given at the moment each streamer reached $L_\mathrm{s}=18$\,mm.
  (e) The simulated and estimated $G$-values for N($^4$S) versus $\overline{E}_\mathrm{max}$ corresponding to panel (a). The estimated $G$-values from equation (\ref{eq:G-value-estimate}) were corrected with the factor $Q_\mathrm{head}/Q_\mathrm{total}$ from panel (d).}
  \label{fig:N4S-G-comparison-negative}
\end{figure}

\begin{figure*}
  \centering
  \includegraphics[width=1\textwidth]{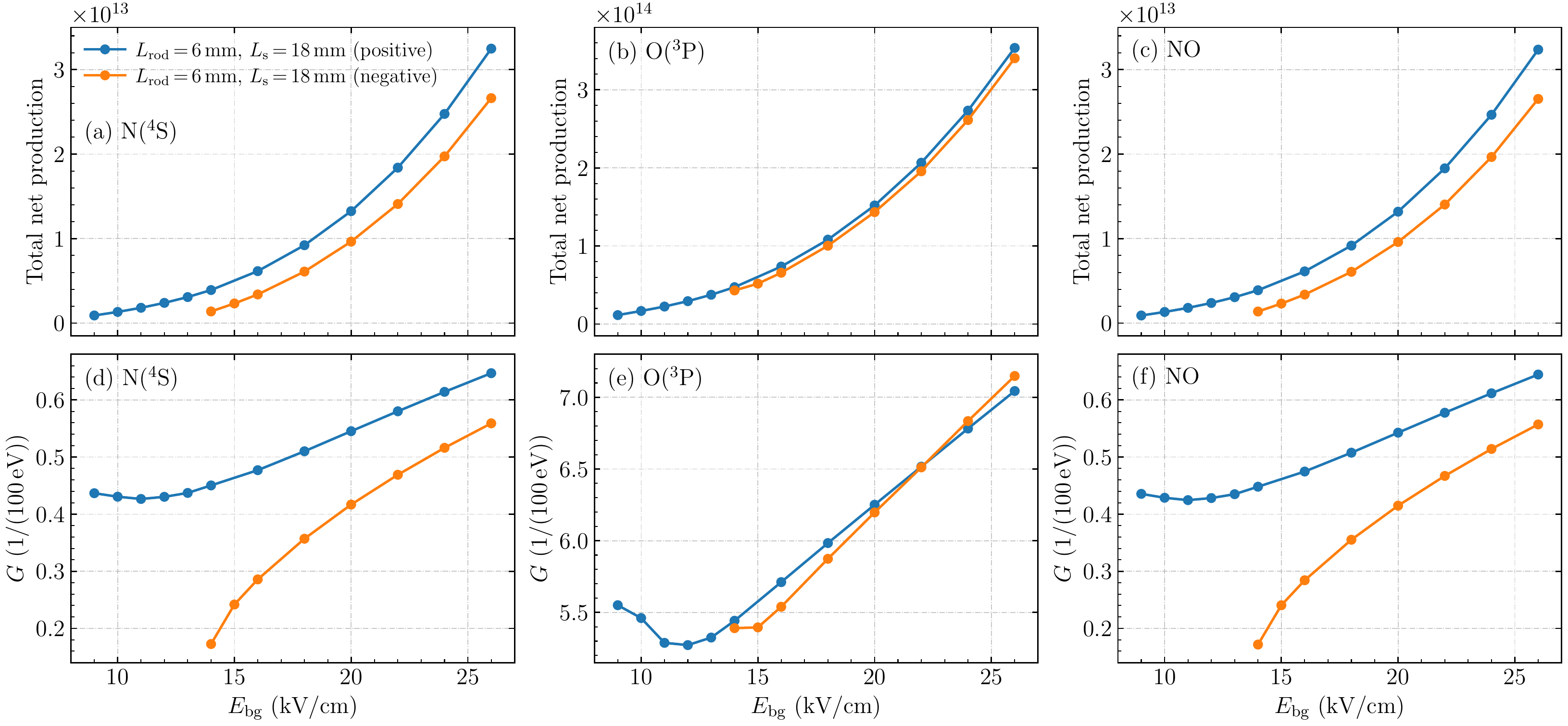}
  \caption{The total net production and $G$-values for N($^4$S), O($^3$P) and NO versus the background field $E_\mathrm{bg}$ for both positive and negative streamers with $L_\mathrm{rod}=6$\,mm, \rev{$R_\mathrm{rod}=$ 0.6\,mm} and $L_\mathrm{s}=18$\,mm.}
  \label{fig:net-production-G-comparison-negative}
\end{figure*}

In the background fields considered here, negative streamers have lower maximal electric fields than positive ones, as shown in figure~\ref{fig:N4S-G-comparison-negative} and also obtained previously~\cite{luque2008, starikovskiy2020}.
Figure~\ref{fig:net-production-G-comparison-negative} shows that total net production and $G$-values for N($^4$S), O($^3$P) and NO are also lower for negative streamers, with the $G$-values for N($^4$S) and NO being up to 60\% lower compared to positive streamers. 

Furthermore, we again compare our results with Naidis' estimate, as discussed in section \ref{sec:naidis-estimate}.
Figure \ref{fig:N4S-G-comparison-negative}(d) shows
that the correction factor $Q_\mathrm{head}/Q_\mathrm{total}$ is smaller for negative streamers, indicating that more energy is deposited in the negative streamer channel.
As shown in figure \ref{fig:N4S-G-comparison-negative}(e), the simulated $G$-values for N($^4$S) agree well with estimated $G$-values for both positive and negative streamers.
For both polarities the same approximately linear relationship between $G$-values and $\overline{E}_\mathrm{max}$ can be observed.

We remark that in~\cite{vanheesch2008}, the authors experimentally measured the energy efficiency of oxygen radical production in primary and secondary streamers for both voltage polarities.
In contrast to our results, they found a higher energy efficiency for negative streamers, but they noted that this efficiency could not accurately be measured for negative streamers due to a larger mismatch of the source impedance.

\subsection{\rev{Qualitative} comparison with experimental data for O$_3$}\label{sec:comparison-experiment}

O$_3$ is produced relatively slowly, on time scales significantly longer than the 500\,ns considered here, as shown in figure \ref{fig:production-vs-time}.
In experiments, O($^3$P) production is usually inferred from the measured O$_3$ production since most (e.g.~85\%) of O($^3$P) is converted to O$_3$ by $\rm O(^3P) + O_2 + M \to O_3 + M$.
\rev{In our simulations} we can therefore estimate $G$-values for O$_3$ based on the O($^3$P) production, which gives $G$-values of about \rev{5--7}, see figure~\ref{fig:nine-G-net-vs-Ebg}(e).


\rev{We can compare these simulated $G$-values to experimentally obtained ones.
However, there are many differences between our simulations and typical experiments, for example in the pulse repetition rate, pulse width, humidity, streamer branching and electrode geometry.
The comparison below is therefore purely qualitative, to check whether the simulated $G$-values are roughly comparable to those in experiments.}

In~\cite{ono2011}, the authors found $G$-values for O$_3$ production in O$_2$(20\%)/N$_2$ of about 4--6, using a 1\,Hz pulse repetition rate in a point-plane gap.
Furthermore, it was observed that the energy efficiency of O$_3$ production was enhanced by reducing the pulse width, and that production was more efficient in primary streamers than in secondary streamers.
In~\cite{winands2006a}, $G$-values for O$_3$ in air of about 2.5--3.4 were found, in a large scale wire-plate reactor with a relative humidity ranging from 35\% to 42\% and pulse repetition rates of 10--400\,Hz.
It was observed that short pulses increased energy efficiencies of O$_3$ production.
In~\cite{wang2010}, a higher $G$-value of about 13 was measured for O$_3$ in dry air at a pulse repetition rate of 30\,Hz in a wire-to-cylinder electrode reactor, using a short voltage pulse of 5\,ns.
In~\cite{vanheesch2008}, the authors proposed that primary streamers were more efficient than secondary streamers for O$_3$ production, and $G$-values of about 10--16 were found for primary positive streamers in air in a wire-plate geometry with pulse repetition rates of 50--100\,Hz, and relative humidity of 13--18\%.

\rev{In conclusion, the simulated $G$-values agree within a factor of two to three with the listed experimental data, which seems reasonable given the different conditions at which the experiments were performed.}


\section{Conclusions and outlook}\label{sec:con-and-outlook}

We have studied how the plasma chemistry of streamer discharges in dry air varies with streamer properties.
Simulations were performed at 300\,K and 1 bar in a 40\,mm plane-plane gap with a needle protrusion, using a 2D axisymmetric fluid model.
Sixty different positive streamers were obtained by varying the electrode \rev{geometry}, the applied voltage and the duration of the voltage pulse.
The obtained streamers had diameters ranging from about 0.5\,mm to 3.5\,mm, velocities ranging from about $0.3\times10^6$\,m/s to $5.7\times10^6$\,m/s and maximal electric fields ranging from about 120\,kV/cm to 150\,kV/cm.

For species with a relatively high activation energy, such as N($^4$S), O($^3$P), NO and N$_2$O, $G$-values were found to vary by about 30\% to \rev{60\%} between the 60 positive cases.
The most important factors behind this variation were differences in the fraction of energy deposited in the streamer head region and variation in the maximal electric field at the streamer head.
When accounting for both factors, good agreement was obtained with an analytic estimate for $G$-values proposed by Naidis~\cite{naidis2012}.
For comparison, eight negative streamers were also simulated.
They were found to be less energy efficient for the production of N($^4$S), O($^3$P) and NO, mainly due to a lower maximal electric field and because more energy was deposited in the streamer channel.

Our results suggest two main ways to increase the energy efficiency with which species with a high activation energy are produced.
First, Joule heating losses in the streamer channel can be reduced, and second, the maximal reduced electric field at the streamer head can be increased.
Short voltage pulses with a high applied voltage can contribute to both factors.

\textbf{Outlook.}
In future work, it would be interesting to study the effect of repetitive pulses, as well as gas heating and humidity effects, and to directly compare with experimental data.
Another question is whether electric fields at streamer heads can be increased by reducing photoionization (for example through the addition of CO$_2$~\cite{bagheri2020}), as this leads to more frequent branching and narrower channels.
Furthermore, it would be interesting to study whether ``minimal streamers'' in low background fields can achieve a high energy efficiency for species production, due to their high maximal electric field~\cite{francisco2021e}.

\rev{Finally, we think testing and improving the validity of the reaction set is of high importance.
This requires careful comparisons against experiments, in which techniques such as optical emission spectroscopy, mass spectrometry or laser-based diagnostics could be used to measure concentrations of neutral, excited and ionized species, as reviewed in e.g.~\cite{simek2014, nijdam2020a}.}

\ack
B.G. was funded by the China Scholarship Council (CSC) (Grant No.\,201906280436).
We thank Dr. Atsushi Komuro from the University of Tokyo for sharing chemical reactions, and we thank Prof. Dr. Ute Ebert, Dr. Bastiaan Braams and Mr. Hemaditya Malla at CWI for their suggestions for improvements.

\section*{Data availability statement}
The data that support the findings of this study are openly available at the following URL/DOI: \url{https://doi.org/10.5281/zenodo.7093075}.
\rev{Any feedback on the reaction set is welcome.}

\appendix


\section{\rev{The reaction set}}
\label{sec:chem-reactions}

Table \ref{tab:electronic-states} lists the excited states of N$_2$ and O$_2$ included in the model, as well as their activation energies.
\rev{The 263 chemical reactions used in the simulations are given} in table~\ref{tab:chem-reactions}, with their rate coefficients and references.
This reaction set contains different processes: rotational excitation, vibrational excitation, electron excitation, electron dissociation, electron impact ionization, electron attachment, electron detachment, negative ion conversion, positive ion conversion, electron-ion recombination, ion-ion recombination and neutral species conversion.

\subsection{\rev{Comments on the reaction set}}
\label{sec:reaction-set-comment}

\rev{The reaction set was constructed for streamer simulations in dry air at 300\,K and 1 bar, considering relatively short time scales (up to several microseconds).
Below we give some comments on the reaction set.

\begin{itemize}
  \item Some of the reaction rate coefficients have to be adjusted when going to different pressures.
  \item Only reactions between electrons and ground-state molecules are taken into account.
  \item No reactions are included to describe the evolution of vibrationally excited states.
  Such reactions are for example given in~\cite{komuro2015a}.
  \item We have not tried to consistently include loss reactions (or reverse reactions) for all species.
  \item There are in particular no loss reactions for N$_2(J)$, N$_2(v)$, O$_2(J)$ and O$_2(v)$.
  When considering secondary streamers, N$_2(v)$ dissociation by electrons can play a role in nitrogen radical formation~\cite{komuro2012}.
  \item High vibrationally excited states N$_2(v > 8)$ and O$_2(v > 4)$ are ignored.
  Such states of N$_2$ can for example contribute to nitrogen radical formation~\cite{naidis2012}.
  \item We have not separated N$_2$(A) into N$_2$(A$_1$) and N$_2$(A$_2$), as was done in e.g.~\cite{fresnet2002, ono2020}. 
  \item Reactions R36 and R201 are actually three-body reactions. These reactions are included as two-body processes because the intermediate states have long lifetimes~\cite{kossyi1992}.
  \item The rate coefficient for reaction R73 is given as $k < 1\times10^{-12} \, \mathrm{cm^3\,s^{-1}}$ in~\cite{kossyi1992}. The upper limit ($k = 1\times10^{-12}$) is used here.
  \item The three-body reactions R89, R97 and R104 include M (any molecule) as a third body.
  Note that in~\cite{kossyi1992} only specific reactions with N$_2$ or O$_2$ are given. However, based on~\cite{aleksandrov1999, guthrie1991} we think using a generic third body is better.
  The rate coefficients for reactions R89 and R104 are taken from~\cite{aleksandrov1999}, and for reaction R97 taken from~\cite{kossyi1992}.
  \item For reaction R121 we assume the `N' from~\cite{kossyi1992} refers to N($^4$S). However, more recent work~\cite{florescumitchell2006} suggests that this reaction produces N($^2$D).
  \item The two-body dissociative reaction of $e$ with O$_2^+$ can produce O($^3$P) and O($^1$D) through different channels. Based on table 30 of~\cite{florescumitchell2006}, the production of O($^3$P) and O($^1$D) is approximately the same.
  For simplicity, we therefore use only one channel producing O($^3$P) + O($^1$D) in reaction R125.
  \item For reactions R157, R159 and R161 from~\cite{gordillo-vazquez2008}, there might be an ``overlap'' with reactions R156, R158 and R160~\cite{kossyi1992}.
  \item Regarding reactions R163 and R164: a total reaction rate coefficient for producing O$_2$(a) or O$_2$(b) is given in~\cite{kossyi1992}. From this total rate coefficient we subtracted the rate coefficient for reaction R164~\cite{popov2011} to obtain reaction R163.
  \item Reactions R196, R234 and R235 from~\cite{kossyi1992} can produce excited states, but due to a lack of data we assume only molecules in the ground state are produced.
\end{itemize}
}

\subsection{\rev{Example of an invalid reaction rate coefficient}}
\label{sec:rates-uncertainty}

\rev{When preparing the reaction set, we also encountered reactions that we later had to exclude.
One such example is the reaction $\rm O_4^+ + O^- \to O_2 + O_3$ from~\cite{tochikubo2002} (2002), with a rate coefficient $$k = 7.80\times10^{-6} \, \mathrm{cm^3\,s^{-1}},$$ citing earlier work~\cite{matzing1991} (1991).
In~\cite{matzing1991}, the rate coefficient is actually given as $$k = 4.0\times10^{-7} (300/T)^{0.5} + 3.0\times 10^{-25} (300/T)^{2.5} M \, \mathrm{cm^3\,s^{-1}},$$ where $T$ is the gas temperature in Kelvin and $M$ is the neutral gas number density, citing~\cite{sutherland1975} (1975).
In~\cite{sutherland1975}, the same rate coefficient can be found (but in a different notation), referring to the third revision of the Defense Nuclear Agency (DNA) Reaction Rate Handbook~\cite{DNA1973} (1973).
We could not obtain the third revision, but in the seventh revision of the DNA handbook, a general recombination reaction of the form
$$\mathrm{X}^+ + \mathrm{Y}^- \to \textrm{products},$$
is given, with a rate coefficient
$$k = 1.0 \times 10^{-7} (300/T)^{0.5},$$
with an uncertainty interval of $(1.0 - 0.6)\times 10^{-7}$ to $(1.0 + 4.0)\times 10^{-7}$.
This last rate coefficient is almost two orders of magnitude lower than the one from~\cite{tochikubo2002}, but it is consistent with~\cite{kossyi1992} and most other literature.
We can therefore conclude that the coefficients given in~\cite{tochikubo2002,matzing1991,sutherland1975} for this recombination reaction (and other similar ones) are likely wrong.
}



\begin{table}
\small
\centering
\caption{Excited states of N$_2$ and O$_2$ with activation energies and the corresponding effective states used in the model. The table is partially based on the one in~\cite{ono2020}.}
\label{tab:electronic-states}
\lineup
\begin{tabular*}{0.49\textwidth}{l@{\extracolsep{\fill}}ccl}
 \br
 \multirow{2}{*}{Excited state} & Activation & \multirow{2}{*}{Effective state} \\ & energy $\epsilon_e$~(eV) \\ 
 \mr
 N$_2$(rot) & 0.02 & N$_2(J)$ \\
 N$_2$($X$, $v=1$) & 0.29 & N$_2(v_1)$ \\
 N$_2$($X$, $v=2$) & 0.59 & N$_2(v_2)$ \\
 N$_2$($X$, $v=3$) & 0.88 & N$_2(v_3)$ \\
 N$_2$($X$, $v=4$) & 1.17 & N$_2(v_4)$ \\
 N$_2$($X$, $v=5$) & 1.47 & N$_2(v_5)$ \\
 N$_2$($X$, $v=6$) & 1.76 & N$_2(v_6)$ \\
 N$_2$($X$, $v=7$) & 2.06 & N$_2(v_7)$ \\
 N$_2$($X$, $v=8$) & 2.35 & N$_2(v_8)$ \\
 N$_2$($A^3\Sigma_u^+$, $v = 0$...4) & 6.17 & \rev{N$_2$(A)} \\
 N$_2$($A^3\Sigma_u^+$, $v = 5$...9) & 7.00 & \rev{N$_2$(A)} \\
 N$_2$($B^3\Pi_g$) & 7.35 & N$_2$(B) \\
 N$_2$($W^3\Delta_u$) & 7.36 & N$_2$(B) \\
 N$_2$($A^3\Sigma_u^+$, $v>10$) & 7.80 & N$_2$(B) \\
 N$_2$($B'^3\Sigma_u^-$) & 8.16 & N$_2$(B) \\
 N$_2$($a'^1\Sigma_u^-$) & 8.40 & N$_2$(a) \\
 N$_2$($a^1\Pi_g$) & 8.55 & N$_2$(a) \\
 N$_2$($w^1\Delta_u$) & 8.89 & N$_2$(a) \\
 N$_2$($C^3\Pi_u$) & 11.03 & N$_2$(C) \\
 N$_2$($E^3\Sigma_g^+$) & 11.87 & N$_2$(E) \\
 N$_2$($a''^1\Sigma_g^+$) & 12.25 & N$_2$(E) \\
 O$_2$(rot) & 0.02 & O$_2(J)$ \\
 O$_2$($X$, $v=1$) & 0.19 & O$_2(v_1)$ \\
 O$_2$($X$, $v=2$) & 0.38 & O$_2(v_2)$ \\
 O$_2$($X$, $v=3$) & 0.57 & O$_2(v_3)$ \\
 O$_2$($X$, $v=4$) & 0.75 & O$_2(v_4)$ \\
 O$_2$($a^1\Delta_g$) & 0.977 & O$_2$(a) \\
 O$_2$($b^1\Sigma_g^+$) & 1.627 & O$_2$(b) \\
 O$_2$($c^1\Sigma_u^-$) & 4.05 & O$_2$(A) \\ 
 O$_2$($A'^3\Delta_u$) & 4.26 & O$_2$(A) \\
 O$_2$($A^3\Sigma_u^+$) & 4.34 & O$_2$(A) \\ 
 \br
\end{tabular*}
\end{table}



\newcounter{nombre}
\renewcommand{\thenombre}{\arabic{nombre}}
\setcounter{nombre}{0}
\newcounter{nombresub} 
\renewcommand{\thenombresub}{\arabic{nombresub}}
\setcounter{nombresub}{0}
\newcommand{\Rnum}[1][]{\refstepcounter{nombre}#1R\thenombre}
\newcommand{\fnum}[1][]{\refstepcounter{nombresub}#1$f_{\thenombresub}$($E/N$)}

\begin{table*}
\renewcommand{\baselinestretch}{1.1}
\footnotesize
\centering
\captionsetup{width=0.95\textwidth}
\caption{List of reactions included in the model, with reaction rate coefficients and references. 
The symbol M denotes a neutral molecule (either $\mathrm{N_2}$ or $\mathrm{O_2}$).
Reaction rate coefficients are in units of $\mathrm{cm^3\,s^{-1}}$ and $\mathrm{cm^6\,s^{-1}}$ for two-body and three-body reactions, respectively.
Reaction rate coefficients for reactions \rev{R1--R30} are functions of the reduced electric field $E/N$, and they are computed with BOLSIG+~\cite{hagelaar2005}. 
$T$(K) and $T_e$(K) are gas and electron temperatures, respectively. 
$T_e(\mathrm{eV})$ is also a measure of electron temperature but in a unit of eV, by multiplying $T_e(\mathrm{K})$ with the Boltzmann constant.}
\label{tab:chem-reactions}
\lineup
\begin{tabular*}{0.95\textwidth}{l@{\extracolsep{\fill}}llll}
 \br
 No. & Reaction & Reaction rate coefficient~($\mathrm{cm^3\,s^{-1}}$ or $\mathrm{cm^6\,s^{-1}}$) & Reference \\ 
 \mr
 \multicolumn{4}{l}{(1) Rotational excitation} \\
 \Rnum & $\rm e + N_2 \to e + N_2({\it J})$ & \fnum & \cite{phelps_database, phelps1985} \\
 \Rnum & $\rm e + O_2 \to e + O_2({\it J})$ & \fnum & \cite{phelps_database, lawton1978} \\
 \mr
 \multicolumn{4}{l}{(2) Vibrational excitation} \\ 
 \Rnum & $\rm e + N_2 \to e + N_2(\it{v}_\mathrm{1})$ & \fnum & \cite{phelps_database, phelps1985} \\
 \Rnum & $\rm e + N_2 \to e + N_2(\it{v}_\mathrm{2})$ & \fnum & \cite{phelps_database, phelps1985} \\
 \Rnum & $\rm e + N_2 \to e + N_2(\it{v}_\mathrm{3})$ & \fnum & \cite{phelps_database, phelps1985} \\
 \Rnum & $\rm e + N_2 \to e + N_2(\it{v}_\mathrm{4})$ & \fnum & \cite{phelps_database, phelps1985} \\
 \Rnum & $\rm e + N_2 \to e + N_2(\it{v}_\mathrm{5})$ & \fnum & \cite{phelps_database, phelps1985} \\
 \Rnum & $\rm e + N_2 \to e + N_2(\it{v}_\mathrm{6})$ & \fnum & \cite{phelps_database, phelps1985} \\
 \Rnum & $\rm e + N_2 \to e + N_2(\it{v}_\mathrm{7})$ & \fnum & \cite{phelps_database, phelps1985} \\
 \Rnum & $\rm e + N_2 \to e + N_2(\it{v}_\mathrm{8})$ & \fnum & \cite{phelps_database, phelps1985} \\
 \Rnum & $\rm e + O_2 \to e + O_2({\it{v}_\mathrm{1}})$ & \fnum & \cite{phelps_database, lawton1978} \\
 \Rnum & $\rm e + O_2 \to e + O_2({\it{v}_\mathrm{2}})$ & \fnum & \cite{phelps_database, lawton1978} \\
 \Rnum & $\rm e + O_2 \to e + O_2({\it{v}_\mathrm{3}})$ & \fnum & \cite{phelps_database, lawton1978} \\
 \Rnum & $\rm e + O_2 \to e + O_2({\it{v}_\mathrm{4}})$ & \fnum & \cite{phelps_database, lawton1978} \\
 \mr
 \multicolumn{4}{l}{(3) Electron excitation} \\ 
 \Rnum & $\rm e + N_2 \to e + N_2(A)$ & \fnum & \cite{phelps_database, phelps1985} \\
 \Rnum & $\rm e + N_2 \to e + N_2(B)$ & \fnum & \cite{phelps_database, phelps1985} \\
 \Rnum & $\rm e + N_2 \to e + N_2(a)$ & \fnum & \cite{phelps_database, phelps1985} \\
 \Rnum & $\rm e + N_2 \to e + N_2(C)$ & \fnum & \cite{phelps_database, phelps1985} \\
 \Rnum & $\rm e + N_2 \to e + N_2(E)$ & \fnum & \cite{phelps_database, phelps1985} \\
 \Rnum & $\rm e + O_2 \to e + O_2(a)$ & \fnum & \cite{phelps_database, lawton1978} \\
 \Rnum & $\rm e + O_2 \to e + O_2(b)$ & \fnum & \cite{phelps_database, lawton1978} \\
 \Rnum & $\rm e + O_2 \to e + O_2(A)$ & \fnum & \cite{phelps_database, lawton1978} \\
 \mr
 \multicolumn{4}{l}{(4) Electron dissociation} \\
 \Rnum & $\rm e + N_2 \to e + N(^4S) + N(^2D)$ & \fnum & \cite{phelps_database, phelps1985} \\
 \Rnum & $\rm e + O_2 \to e + O(^3P) + O(^3P)$ & \fnum & \cite{phelps_database, lawton1978} \\
 \Rnum & $\rm e + O_2 \to e + O(^3P) + O(^1D)$ & \fnum & \cite{phelps_database, lawton1978} \\
 \Rnum & $\rm e + O_2 \to e + O(^3P) + O(^1S)$ & \fnum & \cite{phelps_database, lawton1978} \\ 
 \mr
 \multicolumn{4}{l}{(5) Electron impact ionization} \\
 \Rnum & $\rm e + N_2 \to 2e + N_2^+$ & \fnum & \cite{phelps_database, phelps1985} \\
 \Rnum & $\rm e + O_2 \to 2e + O_2^+$ & \fnum & \cite{phelps_database, lawton1978} \\ 
 \mr
 \multicolumn{4}{l}{(6) Electron attachment} \\
 \Rnum & $\rm e + O_2 + O_2 \to O_2^- + O_2$ & \fnum & \cite{phelps_database, lawton1978} \\
 \Rnum & $\rm e + O_2 \to O^- + O(^3P)$ & \fnum & \cite{phelps_database, lawton1978} \\ 
 \Rnum & $\rm e + O_2 + N_2 \to O_2^- + N_2$ & $1.10\times10^{-31}$ & \cite{gordillo-vazquez2008} \\
 \Rnum & $\rm e + O_3 + O_2 \to O_3^- + O_2$ & $1.00\times10^{-31}$ & \cite{gordillo-vazquez2008} \\
 \Rnum & $\rm e + O_3 \to O_2^- + O(^3P)$ & $1.00\times10^{-9}$ & \cite{kossyi1992} \\
 \Rnum & $\rm e + O_3 \to O^- + O_2$ & $1.00\times10^{-11}$ & \cite{kossyi1992} \\
 \Rnum & $\rm e + NO + M \to NO^- + M$ & $1.00\times10^{-30}$ & \cite{kossyi1992} \\
 \Rnum & $\rm e + NO_2(\,+\,M) \to NO_2^-(\,+\,M)$ & $3.00\times10^{-11}$ & \cite{kossyi1992} \\ 
 \Rnum & $\rm e + NO_2 \to O^- + NO$ & $1.00\times10^{-11}$ & \cite{kossyi1992} \\
 \Rnum & $\rm e + N_2O +N_2 \to N_2O^- + N_2$ & $(4.72(T_e(\mathrm{eV})+0.412)^2-1.268)\times10^{-31}$ & \cite{kossyi1992} \\ 
 \Rnum & $\rm e + O(^3P) + O_2 \to O_2^- + O(^3P)$ & $1.00\times10^{-31}$ & \cite{kossyi1992} \\
 \Rnum & $\rm e + O(^3P) + O_2 \to O^- + O_2$ & $1.00\times10^{-31}$ & \cite{kossyi1992} \\
 \Rnum & $\rm e + O(^3P) + N_2 \to O^- + N_2$ & $1.00\times10^{-31}$ & \cite{gordillo-vazquez2008} \\
 \br
\end{tabular*}
\end{table*}

\addtocounter{table}{-1}

\begin{table*}
\renewcommand{\baselinestretch}{1.12}
\footnotesize
\centering
\captionsetup{width=0.95\textwidth}
\caption{(Continued from previous page)}
\lineup
\begin{tabular*}{0.95\textwidth}{l@{\extracolsep{\fill}}llll}
 \br
 No. & Reaction & Reaction rate coefficient~($\mathrm{cm^3\,s^{-1}}$ or $\mathrm{cm^6\,s^{-1}}$) & Reference \\ 
 \mr
 \multicolumn{4}{l}{(7) Electron detachment} \\
 \Rnum & $\rm O^- + O_2 \to e + O_3$ & $5.00\times10^{-15}$ & \cite{kossyi1992} \\ 
 \Rnum & $\rm O^- + O_2(a) \to e + O_3$ & $3.00\times10^{-10}$ & \cite{kossyi1992} \\ 
 \Rnum & $\rm O^- + O_2(b) \to e + O_2 + O(^3P)$ & $6.90\times10^{-10}$ & \cite{kossyi1992} \\
 \Rnum & $\rm O^- + O_3 \to e + 2O_2$ & $5.30\times10^{-10}$ & \cite{gordillo-vazquez2008} \\ 
 \Rnum & $\rm O^- + N_2 \to e + N_2O$ & $1.16\times10^{-12}\exp(-(\frac{48.9}{11+E/N})^2)$ & \cite{pancheshnyi2013} \\
 \Rnum & $\rm O^- + N_2(A) \to e + N_2 + O(^3P)$ & $2.20\times10^{-9}$ & \cite{kossyi1992} \\
 \Rnum & $\rm O^- + N_2(B) \to e + N_2 + O(^3P)$ & $1.90\times10^{-9}$ & \cite{kossyi1992} \\
 \Rnum & $\rm O^- + NO \to e + NO_2$ & $2.60\times10^{-10}$ & \cite{kossyi1992} \\ 
 \Rnum & $\rm O^- + O(^3P) \to e + O_2$ & $5.00\times10^{-10}$ & \cite{kossyi1992} \\ 
 \Rnum & $\rm O^- + N(^4S) \to e + NO$ & $2.60\times10^{-10}$ & \cite{kossyi1992} \\
 \Rnum & $\rm O_2^- + M \to e + O_2 + M$ & $1.24\times10^{-11}\exp(-(\frac{179}{8.8+E/N})^2)$ & \cite{pancheshnyi2013} \\
 \Rnum & $\rm O_2^- + O_2(a) \to e + 2O_2$ & $2.00\times10^{-10}$ & \cite{kossyi1992} \\ 
 \Rnum & $\rm O_2^- + O_2(b) \to e + 2O_2$ & $3.60\times10^{-10}$ & \cite{kossyi1992} \\ 
 \Rnum & $\rm O_2^- + N_2(A) \to e + O_2 + N_2$ & $2.10\times10^{-9}$ & \cite{kossyi1992} \\ 
 \Rnum & $\rm O_2^- + N_2(B) \to e + O_2 + N_2$ & $2.50\times10^{-9}$ & \cite{kossyi1992} \\
 \Rnum & $\rm O_2^- + O(^3P) \to e + O_3$ & $1.50\times10^{-10}$ & \cite{kossyi1992} \\ 
 \Rnum & $\rm O_2^- + N(^4S) \to e + NO_2$ & $5.00\times10^{-10}$ & \cite{kossyi1992} \\ 
 \Rnum & $\rm O_3^- + O_3 \to e + 3O_2$ & $1.00\times10^{-10}$ & \cite{gordillo-vazquez2008} \\
 \Rnum & $\rm O_3^- + O(^3P) \to e + 2O_2$ & $3.00\times10^{-10}$ & \cite{kossyi1992} \\ 
 \mr
 \multicolumn{4}{l}{(8) Negative ion conversion} \\
 \Rnum & $\rm O^- + O_2 + M \to O_3^- + M$ & $1.10\times10^{-30}\exp(-(\frac{E/N}{65})^2)$ & \cite{pancheshnyi2013} \\ 
 \Rnum & $\rm O^- + O_2 \to O_2^- + O(^3P)$ & $6.96\times10^{-11}\exp(-(\frac{198}{5.6+E/N})^2)$ & \cite{pancheshnyi2013} \\ 
 \Rnum & $\rm O^- + O_2(a) \to O_2^- + O(^3P)$ & $1.00\times10^{-10}$ & \cite{kossyi1992} \\
 \Rnum & $\rm O^- + O_3 \to O_3^- + O(^3P)$ & $5.30\times10^{-10}$ & \cite{kossyi1992} \\ 
 \Rnum & $\rm O^- + NO + M \to NO_2^- + M$ & $1.00\times10^{-29}$ & \cite{kossyi1992} \\
 \Rnum & $\rm O^- + NO_2 \to NO_2^- + O(^3P)$ & $1.20\times10^{-9}$ & \cite{kossyi1992} \\ 
 \Rnum & $\rm O^- + N_2O \to NO^- + NO$ & $2.00\times10^{-10}$ & \cite{kossyi1992} \\ 
 \Rnum & $\rm O^- + N_2O \to N_2O^- + O(^3P)$ & $2.00\times10^{-12}$ & \cite{kossyi1992} \\ 
 \Rnum & $\rm O_2^- + O_2 + M \to O_4^- + M$ & $3.50\times10^{-31}(\frac{300}{T})$ & \cite{kossyi1992} \\
 \Rnum & $\rm O_2^- + O_3 \to O_3^- + O_2$ & $4.00\times10^{-10}$ & \cite{kossyi1992} \\ 
 \Rnum & $\rm O_2^- + NO_2 \to NO_2^- + O_2$ & $8.00\times10^{-10}$ & \cite{kossyi1992} \\ 
 \Rnum & $\rm O_2^- + NO_3 \to NO_3^- + O_2$ & $5.00\times10^{-10}$ & \cite{kossyi1992} \\ 
 \Rnum & $\rm O_2^- + N_2O \to O_3^- + N_2$ & $1.00\times10^{-12}$ & \cite{kossyi1992} \\ 
 \Rnum & $\rm O_2^- + O(^3P) \to O^- + O_2$ & $3.30\times10^{-10}$ & \cite{kossyi1992} \\
 \Rnum & $\rm O_3^- + NO \to NO_2^- + O_2$ & $2.60\times10^{-12}$ & \cite{kossyi1992} \\
 \Rnum & $\rm O_3^- + NO \to NO_3^- + O(^3P)$ & $1.00\times10^{-11}$ & \cite{kossyi1992} \\
 \Rnum & $\rm O_3^- + NO_2 \to NO_2^- + O_3$ & $7.00\times10^{-10}$ & \cite{kossyi1992} \\
 \Rnum & $\rm O_3^- + NO_2 \to NO_3^- + O_2$ & $2.00\times10^{-11}$ & \cite{kossyi1992} \\
 \Rnum & $\rm O_3^- + NO_3 \to NO_3^- + O_3$ & $5.00\times10^{-10}$ & \cite{kossyi1992} \\
 \Rnum & $\rm O_3^- + O(^3P) \to O_2^- + O_2$ & $3.20\times10^{-10}$ & \cite{kossyi1992} \\
 \Rnum & $\rm O_4^- + M \to O_2^- + O_2 + M$ & $1.00\times10^{-10}\exp(\frac{-1044}{T})$ & \cite{kossyi1992} \\
 \Rnum & $\rm O_4^- + O_2(a) \to O_2^- + 2O_2$ & $1.00\times10^{-10}$ & \cite{kossyi1992} \\
 \Rnum & $\rm O_4^- + O_2(b) \to O_2^- + 2O_2$ & $1.00\times10^{-10}$ & \cite{kossyi1992} \\
 \Rnum & $\rm O_4^- + NO \to NO_3^- + O_2$ & $2.50\times10^{-10}$ & \cite{kossyi1992} \\
 \Rnum & $\rm O_4^- + O(^3P) \to O^- + 2O_2$ & $3.00\times10^{-10}$ & \cite{kossyi1992} \\
 \Rnum & $\rm O_4^- + O(^3P) \to O_3^- + O_2$ & $4.00\times10^{-10}$ & \cite{kossyi1992} \\ 
 \mr
 \multicolumn{4}{l}{(9) Positive ion conversion} \\
 \Rnum & $\rm N_2^+ + O_2 \to O_2^+ + N_2$ & $6.00\times10^{-11}(\frac{300}{T})^{0.5}$ & \cite{kossyi1992} \\
 \Rnum & $\rm N_2^+ + O_3 \to O_2^+ + N_2 + O(^3P)$ & $1.00\times10^{-10}$ & \cite{kossyi1992} \\
 \Rnum & $\rm N_2^+ + N_2 + M \to N_4^+ + M$ & $5.00\times10^{-29}(\frac{300}{T})^{2}$ & \cite{kossyi1992, aleksandrov1999} \\
 \Rnum & $\rm N_2^+ + N_2(A) \to N_3^+ + N(^4S)$ & $3.00\times10^{-10}$ & \cite{kossyi1992} \\
 \Rnum & $\rm N_2^+ + NO \to NO^+ + N_2$ & $3.30\times10^{-10}$ & \cite{kossyi1992} \\
 \Rnum & $\rm N_2^+ + NO_2 \to NO_2^+ + N_2$ & $3.00\times10^{-10}$ & \cite{tochikubo2002} \\
 \Rnum & $\rm N_2^+ + N_2O \to NO^+ + N_2 + N(^4S)$ & $4.00\times10^{-10}$ & \cite{kossyi1992} \\
 \Rnum & $\rm N_2^+ + N_2O \to N_2O^+ + N_2$ & $5.00\times10^{-10}$ & \cite{kossyi1992} \\
 \br
\end{tabular*}
\end{table*}

\addtocounter{table}{-1}

\begin{table*}
\renewcommand{\baselinestretch}{1.12}
\footnotesize
\centering
\captionsetup{width=0.95\textwidth}
\caption{(Continued from previous page)}
\lineup
\begin{tabular*}{0.95\textwidth}{l@{\extracolsep{\fill}}llll}
 \br
 No. & Reaction & Reaction rate coefficient~($\mathrm{cm^3\,s^{-1}}$ or $\mathrm{cm^6\,s^{-1}}$) & Reference \\ 
 \mr
 \Rnum & $\rm N_2^+ + O(^3P) \to O^+ + N_2$ & $1.00\times10^{-11}(\frac{300}{T})^{0.2}$ & \cite{kossyi1992} \\
 \Rnum & $\rm N_2^+ + O(^3P) \to NO^+ + N(^4S)$ & $1.30\times10^{-10}(\frac{300}{T})^{0.5}$ & \cite{kossyi1992} \\
 \Rnum & $\rm N_2^+ + N(^4S) + M \to N_3^+ + M$ & $9.00\times10^{-30}\exp(\frac{400}{T})$ & \cite{kossyi1992, aleksandrov1999} \\
 \Rnum & $\rm N_2^+ + N(^4S) \to N^+ + N_2$ & $2.40\times10^{-15}T$ & \cite{kossyi1992} \\
 \Rnum & $\rm N_4^+ + O_2 \to O_2^+ + 2N_2$ & $2.50\times10^{-10}$ & \cite{kossyi1992} \\
 \Rnum & $\rm N_4^+ + N_2 \to N_2^+ + 2N_2$ & $10^{-14.6+0.0036(T-300)}$ & \cite{kossyi1992} \\
 \Rnum & $\rm N_4^+ + NO \to NO^+ + 2N_2$ & $4.00\times10^{-10}$ & \cite{kossyi1992} \\
 \Rnum & $\rm N_4^+ + O(^3P) \to O^+ + 2N_2$ & $2.50\times10^{-10}$ & \cite{kossyi1992} \\
 \Rnum & $\rm N_4^+ + N(^4S) \to N^+ + 2N_2$ & $1.00\times10^{-11}$ & \cite{kossyi1992} \\
 \Rnum & $\rm O_2^+ + O_2 + M \to O_4^+ + M$ & $2.40\times10^{-30}(\frac{300}{T})^{3}$ & \cite{kossyi1992, aleksandrov1999} \\
 \Rnum & $\rm O_2^+ + N_2 + N_2 \to N_2O_2^+ + N_2$ & $9.00\times10^{-31}(\frac{300}{T})^{2}$ & \cite{kossyi1992} \\ 
 \Rnum & $\rm O_2^+ + N_2 \to NO^+ + NO$ & $1.00\times10^{-17}$ & \cite{kossyi1992} \\
 \Rnum & $\rm O_2^+ + NO \to NO^+ + O_2$ & $4.40\times10^{-10}$ & \cite{kossyi1992} \\
 \Rnum & $\rm O_2^+ + NO_2 \to NO^+ + O_3$ & $1.00\times10^{-11}$ & \cite{kossyi1992} \\
 \Rnum & $\rm O_2^+ + NO_2 \to NO_2^+ + O_2$ & $6.60\times10^{-10}$ & \cite{kossyi1992} \\
 \Rnum & $\rm O_2^+ + N(^4S) \to NO^+ + O(^3P)$ & $1.20\times10^{-10}$ & \cite{kossyi1992} \\
 \Rnum & $\rm O_4^+ + O_2 \to O_2^+ + 2O_2$ & $3.30\times10^{-6}(\frac{300}{T})^{4}\exp(\frac{-5030}{T})$ & \cite{kossyi1992} \\
 \Rnum & $\rm O_4^+ + O_2(a) \to O_2^+ + 2O_2$ & $1.00\times10^{-10}$ & \cite{kossyi1992} \\
 \Rnum & $\rm O_4^+ + O_2(b) \to O_2^+ + 2O_2$ & $1.00\times10^{-10}$ & \cite{kossyi1992} \\
 \Rnum & $\rm O_4^+ + N_2 \to N_2O_2^+ + O_2$ & $4.61\times10^{-12}(\frac{T}{300})^{2.5}\exp(\frac{-2650}{T})$ & \cite{kossyi1992} \\
 \Rnum & $\rm O_4^+ + NO \to NO^+ + 2O_2$ & $1.00\times10^{-10}$ & \cite{kossyi1992} \\
 \Rnum & $\rm O_4^+ + NO_2 \to NO_2^+ + 2O_2$ & $3.00\times10^{-10}$ & \cite{tochikubo2002} \\
 \Rnum & $\rm O_4^+ + O(^3P) \to O_2^+ + O_3$ & $3.00\times10^{-10}$ & \cite{kossyi1992} \\
 \Rnum & $\rm N_2O_2^+ + O_2 \to O_4^+ + N_2$ & $1.00\times10^{-9}$ & \cite{kossyi1992} \\
 \Rnum & $\rm N_2O_2^+ + N_2 \to O_2^+ + 2N_2$ & $1.10\times10^{-6}(\frac{300}{T})^{5.3}\exp(\frac{-2357}{T})$ & \cite{kossyi1992} \\ 
 \mr
 \multicolumn{4}{l}{(10) Electron-ion recombination} \\
 \Rnum & $\rm e + N_2^+ + M \to N_2 + M$ & $6.00\times10^{-27}(\frac{300}{T_e})^{1.5}$ & \cite{kossyi1992} \\
 \Rnum & $\rm e + N_2^+ \to N(^4S) + N(^4S)$ & $2.80\times10^{-7}(\frac{300}{T_e})^{0.5}$ & \cite{kossyi1992, florescumitchell2006} \\
 \Rnum & $\rm e + N_2^+ \to N(^4S) + N(^2D)$ & $2.00\times10^{-7}(\frac{300}{T_e})^{0.5}$ & \cite{kossyi1992} \\
 \Rnum & $\rm e + N_4^+ \to 2N_2$ & $2.00\times10^{-6}(\frac{300}{T_e})^{0.5}$ & \cite{kossyi1992} \\
 \Rnum & $\rm e + O_2^+ + M \to O_2 + M$ & $6.00\times10^{-27}(\frac{300}{T_e})^{1.5}$ & \cite{kossyi1992} \\
 \Rnum & $\rm e + O_2^+ \to O(^3P) + O(^1D)$ & $1.95\times10^{-7}(\frac{300}{T_e})^{0.7}$ & \cite{kossyi1992, florescumitchell2006} \\
 \Rnum & $\rm e + O_4^+ \to 2O_2$ & $1.40\times10^{-6}(\frac{300}{T_e})^{0.5}$ & \cite{kossyi1992} \\
 \Rnum & $\rm e + N_2O_2^+ \to N_2 + O_2$ & $1.30\times10^{-6}(\frac{300}{T_e})^{0.5}$ & \cite{kossyi1992} \\ 
 \mr
 \multicolumn{4}{l}{(11) Ion-ion recombination} \\
 \Rnum & $\rm N_2^+ + O^- + M \to N_2O + M$ & $2.00\times10^{-25}(\frac{300}{T})^{2.5}$ & \cite{kossyi1992} \\
 \Rnum & $\rm N_2^+ + O^- + M \to N_2 + O(^3P) + M$ & $2.00\times10^{-25}(\frac{300}{T})^{2.5}$ & \cite{kossyi1992} \\
 \Rnum & $\rm N_2^+ + O^- \to N_2 + O(^3P)$ & $2.00\times10^{-7}(\frac{300}{T})^{0.5}$ & \cite{kossyi1992} \\
 \Rnum & $\rm N_2^+ + O^- \to 2N(^4S) + O(^3P)$ & $1.00\times10^{-7}$ & \cite{kossyi1992} \\
 \Rnum & $\rm N_2^+ + O_2^- + M \to N_2 + O_2 + M$ & $2.00\times10^{-25}(\frac{300}{T})^{2.5}$ & \cite{kossyi1992} \\
 \Rnum & $\rm N_2^+ + O_2^- \to N_2 + O_2$ & $2.00\times10^{-7}(\frac{300}{T})^{0.5}$ & \cite{kossyi1992} \\
 \Rnum & $\rm N_2^+ + O_2^- \to O_2 + 2N(^4S)$ & $1.00\times10^{-7}$ & \cite{kossyi1992} \\
 \Rnum & $\rm N_2^+ + O_3^- \to N_2 + O_3$ & $2.00\times10^{-7}(\frac{300}{T})^{0.5}$ & \cite{kossyi1992} \\
 \Rnum & $\rm N_2^+ + O_3^- \to O_3 + 2N(^4S)$ & $1.00\times10^{-7}$ & \cite{kossyi1992} \\
 \Rnum & $\rm N_2^+ + O_4^- \to N_2 + 2O_2$ & $1.00\times10^{-7}$ & \cite{kossyi1992} \\
 \Rnum & $\rm N_4^+ + O^- \to 2N_2 + O(^3P)$ & $1.00\times10^{-7}$ & \cite{kossyi1992} \\
 \Rnum & $\rm N_4^+ + O_2^- \to 2N_2 + O_2$ & $1.00\times10^{-7}$ & \cite{kossyi1992} \\
 \Rnum & $\rm N_4^+ + O_3^- \to 2N_2 + O_3$ & $1.00\times10^{-7}$ & \cite{kossyi1992} \\
 \Rnum & $\rm N_4^+ + O_4^- \to 2N_2 + 2O_2$ & $1.00\times10^{-7}$ & \cite{kossyi1992} \\
 \Rnum & $\rm O_2^+ + O^- + M \to O_3 + M$ & $2.00\times10^{-25}(\frac{300}{T})^{2.5}$ & \cite{kossyi1992} \\
 \Rnum & $\rm O_2^+ + O^- + M \to O_2 + O(^3P) + M$ & $2.00\times10^{-25}(\frac{300}{T})^{2.5}$ & \cite{kossyi1992} \\
 \Rnum & $\rm O_2^+ + O^- \to O_2 + O(^3P)$ & $2.00\times10^{-7}(\frac{300}{T})^{0.5}$ & \cite{kossyi1992} \\
 \Rnum & $\rm O_2^+ + O^- \to 3O(^3P)$ & $1.00\times10^{-7}$ & \cite{kossyi1992} \\
 \Rnum & $\rm O_2^+ + O_2^- + M \to 2O_2 + M$ & $2.00\times10^{-25}(\frac{300}{T})^{2.5}$ & \cite{kossyi1992} \\
 \Rnum & $\rm O_2^+ + O_2^- \to 2O_2$ & $2.00\times10^{-7}(\frac{300}{T})^{0.5}$ & \cite{kossyi1992} \\
 \Rnum & $\rm O_2^+ + O_2^- \to O_2 + 2O(^3P)$ & $1.00\times10^{-7}$ & \cite{kossyi1992} \\
 \br
\end{tabular*}
\end{table*}

\addtocounter{table}{-1}

\begin{table*}
\renewcommand{\baselinestretch}{1.12}
\footnotesize
\centering
\captionsetup{width=0.95\textwidth}
\caption{(Continued from previous page)}
\lineup
\begin{tabular*}{0.95\textwidth}{l@{\extracolsep{\fill}}llll}
 \br
 No. & Reaction & Reaction rate coefficient~($\mathrm{cm^3\,s^{-1}}$ or $\mathrm{cm^6\,s^{-1}}$) & Reference \\ 
 \mr
 \Rnum & $\rm O_2^+ + O_3^- \to O_2 + O_3$ & $2.00\times10^{-7}(\frac{300}{T})^{0.5}$ & \cite{kossyi1992} \\
 \Rnum & $\rm O_2^+ + O_3^- \to O_3 + 2O(^3P)$ & $1.00\times10^{-7}$ & \cite{kossyi1992} \\
 \Rnum & $\rm O_2^+ + O_4^- \to 3O_2$ & $1.00\times10^{-7}$ & \cite{kossyi1992} \\
 \Rnum & $\rm O_4^+ + O^- \to 2O_2 + O(^3P)$ & $1.00\times10^{-7}$ & \cite{kossyi1992} \\
 \Rnum & $\rm O_4^+ + O_2^- \to 3O_2$ & $1.00\times10^{-7}$ & \cite{kossyi1992} \\
 \Rnum & $\rm O_4^+ + O_3^- \to 2O_2 + O_3$ & $1.00\times10^{-7}$ & \cite{kossyi1992} \\
 \Rnum & $\rm O_4^+ + O_4^- \to 4O_2$ & $1.00\times10^{-7}$ & \cite{kossyi1992} \\
 \Rnum & $\rm N_2O_2^+ + O^- \to N_2 + O_2 + O(^3P)$ & $1.00\times10^{-7}$ & \cite{kossyi1992} \\
 \Rnum & $\rm N_2O_2^+ + O^- \to 2NO + O(^3P)$ & $1.00\times10^{-7}$ & \cite{gordillo-vazquez2008} \\
 \Rnum & $\rm N_2O_2^+ + O_2^- \to N_2 + 2O_2$ & $1.00\times10^{-7}$ & \cite{kossyi1992} \\
 \Rnum & $\rm N_2O_2^+ + O_2^- \to 2NO + O_2$ & $1.00\times10^{-7}$ & \cite{gordillo-vazquez2008} \\
 \Rnum & $\rm N_2O_2^+ + O_3^- \to N_2 + O_2 + O_3$ & $1.00\times10^{-7}$ & \cite{kossyi1992} \\
  \Rnum & $\rm N_2O_2^+ + O_3^- \to 2NO + O_3$ & $1.00\times10^{-7}$ & \cite{gordillo-vazquez2008} \\
 \Rnum & $\rm N_2O_2^+ + O_4^- \to N_2 + 3O_2$ & $1.00\times10^{-7}$ & \cite{kossyi1992} \\ 
 \mr
 \multicolumn{4}{l}{(12) Neutral species conversion} \\
 \Rnum & $\rm N_2(A) + O_2 \to N_2 + O_2(a)$ & $5.40\times10^{-13}$ & \cite{kossyi1992, popov2011} \\
 \Rnum & $\rm N_2(A) + O_2 \to N_2 + O_2(b)$ & $7.50\times10^{-13}$ & \cite{popov2011} \\
 \Rnum & $\rm N_2(A) + O_2 \to N_2 + 2O(^3P)$ & $2.54\times10^{-12}$ & \cite{kossyi1992} \\
 \Rnum & $\rm N_2(A) + O_2 \to N_2O + O(^3P)$ & $7.80\times10^{-14}$ & \cite{kossyi1992} \\
 \Rnum & $\rm N_2(A) + N_2 \to 2N_2$ & $3.00\times10^{-18}$ & \cite{kossyi1992} \\
 \Rnum & $\rm N_2(A) + N_2(A) \to N_2 + N_2(B)$ & $7.70\times10^{-11}$ & \cite{popov2011} \\
 \Rnum & $\rm N_2(A) + N_2(A) \to N_2 + N_2(C)$ & $1.60\times10^{-10}$ & \cite{kossyi1992} \\
 \Rnum & $\rm N_2(A) + NO \to N_2 + NO$ & $7.00\times10^{-11}$ & \cite{kossyi1992} \\
 \Rnum & $\rm N_2(A) + N_2O \to N_2 + NO + N(^4S)$ & $1.00\times10^{-11}$ & \cite{kossyi1992} \\
 \Rnum & $\rm N_2(A) + O(^3P) \to N_2 + O(^1S)$ & $2.10\times10^{-11}$ & \cite{kossyi1992} \\
 \Rnum & $\rm N_2(A) + O(^3P) \to NO + N(^2D)$ & $7.00\times10^{-12}$ & \cite{kossyi1992} \\
 \Rnum & $\rm N_2(A) + N(^4S) \to N_2 + N(^4S)$ & $2.00\times10^{-12}$ & \cite{gordiets1995} \\
 \Rnum & $\rm N_2(A) + N(^4S) \to N_2 + N(^2P)$ & $5.00\times10^{-11}$ & \cite{kossyi1992} \\
 \Rnum & $\rm N_2(B) + O_2 \to N_2 + 2O(^3P)$ & $3.00\times10^{-10}$ & \cite{kossyi1992} \\
 \Rnum & $\rm N_2(B) + N_2 \to 2N_2$ & $2.00\times10^{-12}$ & \cite{gordillo-vazquez2008} \\
 \Rnum & $\rm N_2(B) + N_2 \to N_2 + N_2(A)$ & $5.00\times10^{-11}$ & \cite{kossyi1992} \\
 \Rnum & $\rm N_2(B) + NO \to NO + N_2(A)$ & $2.40\times10^{-10}$ & \cite{kossyi1992} \\
 \Rnum & $\rm N_2(B) \to N_2(A) + {\it h\nu}$ & $1.50\times10^{5}$ (s$^{-1}$) & \cite{kossyi1992} \\
 \Rnum & $\rm N_2(a) + O_2 \to N_2 + O(^3P) + O(^3P)$ & $2.80\times10^{-11}$ &\cite{kossyi1992} \\
 \Rnum & $\rm N_2(a) + N_2 \to N_2 + N_2(B)$ & $2.00\times10^{-13}$ & \cite{kossyi1992} \\
 \Rnum & $\rm N_2(a) + NO \to N_2 + N(^4S) + O(^3P)$ & $3.60\times10^{-10}$ & \cite{kossyi1992} \\
 \Rnum & $\rm N_2(C) + O_2 \to N_2 + O(^3P) + O(^1S)$ & $3.00\times10^{-10}$ & \cite{kossyi1992} \\
 \Rnum & $\rm N_2(C) + N_2 \to N_2 + N_2(a)$ & $1.00\times10^{-11}$ & \cite{kossyi1992} \\
 \Rnum & $\rm N_2(C) \to N_2(B) + {\it h\nu}$ & $3.00\times10^{7}$ (s$^{-1}$) & \cite{kossyi1992} \\
 \Rnum & $\rm N_2(E) + N_2 \to N_2 + N_2(C)$ & $1.00\times10^{-11}$ & \cite{fresnet2002} \\
 \Rnum & $\rm N(^4S) + O_2 \to NO + O(^3P)$ & $4.50\times10^{-12}\exp(\frac{-3220}{T})$ & \cite{kossyi1992} \\
 \Rnum & $\rm N(^4S) + O_3 \to NO + O_2$ & $2.00\times10^{-16}$ & \cite{kossyi1992} \\
 \Rnum & $\rm N(^4S) + NO \to N_2 + O(^3P)$ & $1.05\times10^{-12}T^{0.5}$ & \cite{kossyi1992} \\
 \Rnum & $\rm N(^4S) + NO_2 \to N_2 + O_2$ & $7.00\times10^{-13}$ & \cite{kossyi1992} \\
 \Rnum & $\rm N(^4S) + NO_2 \to N_2 + 2O(^3P)$ & $9.10\times10^{-13}$ & \cite{kossyi1992} \\
 \Rnum & $\rm N(^4S) + NO_2 \to 2NO$ & $2.30\times10^{-12}$ & \cite{kossyi1992} \\
 \Rnum & $\rm N(^4S) + NO_2 \to N_2O + O(^3P)$ & $3.00\times10^{-12}$ & \cite{kossyi1992} \\
 \Rnum & $\rm N(^4S) + O(^3P) + M \to NO + M$ & $1.76\times10^{-31}T^{-0.5}$ & \cite{kossyi1992} \\
 \Rnum & $\rm N(^4S) + N(^4S) + M \to N_2 + M$ & $8.27\times10^{-34}\exp(\frac{500}{T})$ & \cite{kossyi1992} \\
 \Rnum & $\rm N(^4S) + N(^2P) \to N(^4S) + N(^2D)$ & $1.80\times10^{-12}$ & \cite{kossyi1992} \\
 \Rnum & $\rm N(^2D) + O_2 \to NO + O(^3P)$ & $1.50\times10^{-12}(\frac{T}{300})^{0.5}$ & \cite{kossyi1992} \\
 \Rnum & $\rm N(^2D) + O_2 \to NO + O(^1D)$ & $6.00\times10^{-12}(\frac{T}{300})^{0.5}$ & \cite{kossyi1992} \\
 \Rnum & $\rm N(^2D) + N_2 \to N_2 + N(^4S)$ & $6.00\times10^{-15}$ & \cite{kossyi1992} \\
 \Rnum & $\rm N(^2D) + NO (\,+\,M) \to N_2O (\,+\,M)$ & $6.00\times10^{-11}$ & \cite{kossyi1992} \\
 \Rnum & $\rm N(^2D) + N_2O \to NO + N_2$ & $3.00\times10^{-12}$ & \cite{kossyi1992} \\
 \Rnum & $\rm N(^2P) + O_2 \to NO + O(^3P)$ & $2.60\times10^{-12}$ & \cite{kossyi1992} \\
 \Rnum & $\rm N(^2P) + N_2 \to N_2 + N(^2D)$ & $2.00\times10^{-18}$ & \cite{kossyi1992} \\
 \Rnum & $\rm N(^2P) + NO \to N_2(A) + O(^3P)$ & $3.40\times10^{-11}$ & \cite{kossyi1992} \\
 \br
\end{tabular*}
\end{table*}

\addtocounter{table}{-1}

\begin{table*}
\renewcommand{\baselinestretch}{1.12}
\footnotesize
\centering
\captionsetup{width=0.95\textwidth}
\caption{(Continued from previous page)}
\lineup
\begin{tabular*}{0.95\textwidth}{l@{\extracolsep{\fill}}llll}
 \br
 No. & Reaction & Reaction rate coefficient~($\mathrm{cm^3\,s^{-1}}$ or $\mathrm{cm^6\,s^{-1}}$) & Reference \\ 
 \mr
 \Rnum & $\rm O_2(a) + O_2 \to 2O_2$ & $2.20\times10^{-18}(\frac{T}{300})^{0.8}$ & \cite{kossyi1992} \\
 \Rnum & $\rm O_2(a) + O_2(a) + O_2 \to 2O_3$ & $1.00\times10^{-31}$ & \cite{gordiets1995} \\
 \Rnum & $\rm O_2(a) + O_2(a) \to O_2 + O_2(b)$ & $7.00\times10^{-28}T^{3.8}\exp(\frac{700}{T})$ & \cite{gordiets1995} \\
 \Rnum & $\rm O_2(a) + O_3 \to 2O_2 + O(^3P)$ & $9.70\times10^{-13}\exp(\frac{-1564}{T})$ & \cite{kossyi1992} \\
 \Rnum & $\rm O_2(a) + N_2 \to O_2 + N_2$ & $3.00\times10^{-21}$ & \cite{kossyi1992} \\
 \Rnum & $\rm O_2(a) + NO \to NO + O_2$ & $2.50\times10^{-11}$ & \cite{kossyi1992} \\
 \Rnum & $\rm O_2(a) + O(^3P) \to O_2 + O(^3P)$ & $7.00\times10^{-16}$ & \cite{kossyi1992} \\
 \Rnum & $\rm O_2(a) + O(^1S) \to 3O(^3P)$ & $3.40\times10^{-11}$ & \cite{kossyi1992} \\
 \Rnum & $\rm O_2(a) + O(^1S) \to O_2(b) + O(^1D)$ & $3.60\times10^{-11}$ & \cite{kossyi1992} \\
 \Rnum & $\rm O_2(a) + O(^1S) \to O_2(A) + O(^3P)$ & $1.30\times10^{-10}$ & \cite{kossyi1992} \\
 \Rnum & $\rm O_2(a) + N(^4S) \to NO + O(^3P)$ & $2.00\times10^{-14}\exp(\frac{-600}{T})$ & \cite{kossyi1992} \\ 
 \Rnum & $\rm O_2(b) + O_2 \to O_2 + O_2(a)$ & $4.30\times10^{-22}T^{2.4}\exp(\frac{-241}{T})$ & \cite{kossyi1992} \\
 \Rnum & $\rm O_2(b) + O_3 \to 2O_2 + O(^3P)$ & $1.80\times10^{-11}$ & \cite{kossyi1992} \\
 \Rnum & $\rm O_2(b) + N_2 \to N_2 + O_2(a)$ & $4.90\times10^{-15}\exp(\frac{-253}{T})$ & \cite{kossyi1992} \\
 \Rnum & $\rm O_2(b) + NO \to NO + O_2(a)$ & $4.00\times10^{-14}$ & \cite{kossyi1992} \\
 \Rnum & $\rm O_2(b) + O(^3P) \to O_2(a) + O(^3P)$ & $8.00\times10^{-14}$ & \cite{kossyi1992} \\
 \Rnum & $\rm O_2(b) + O(^3P) \to O_2 + O(^1D)$ & $3.39\times10^{-11}(\frac{300}{T})^{0.1}\exp(\frac{-4201}{T})$ & \cite{kossyi1992} \\
 \Rnum & $\rm O_2(A) + O_2 \to 2O_2(b)$ & $2.90\times10^{-13}$ & \cite{kossyi1992} \\
 \Rnum & $\rm O_2(A) + N_2 \to N_2 + O_2(b)$ & $3.00\times10^{-13}$ & \cite{kossyi1992} \\
 \Rnum & $\rm O_2(A) + O(^3P) \to O_2(b) + O(^1D)$ & $9.00\times10^{-12}$ & \cite{kossyi1992} \\ 
 \Rnum & $\rm O(^3P) + O_2 + O_2 \to O_3 + O_2$ & $6.90\times10^{-34}(\frac{300}{T})^{1.25}$ & \cite{kossyi1992} \\
 \Rnum & $\rm O(^3P) + O_2 + N_2 \to O_3 + N_2$ & $6.20\times10^{-34}(\frac{300}{T})^{2}$ & \cite{kossyi1992} \\
 \Rnum & $\rm O(^3P) + O_3 \to 2O_2$ & $2.00\times10^{-11}\exp(\frac{-2300}{T})$ & \cite{kossyi1992} \\
 \Rnum & $\rm O(^3P) + NO + O_2 \to NO_2 + O_2$ & $9.30\times10^{-32}(\frac{300}{T})^{1.682}$ & \cite{gordillo-vazquez2008} \\
 \Rnum & $\rm O(^3P) + NO + N_2 \to NO_2 + N_2$ & $1.20\times10^{-31}(\frac{300}{T})^{1.682}$ & \cite{gordillo-vazquez2008} \\
 \Rnum & $\rm O(^3P) + NO_2 + M \to NO_3 + M$ & $8.90\times10^{-32}(\frac{300}{T})^{2}$ & \cite{gordillo-vazquez2008} \\
 \Rnum & $\rm O(^3P) + NO_2 \to NO + O_2$ & $1.13\times10^{-11}(\frac{T}{1000})^{0.18}$ & \cite{kossyi1992} \\
 \Rnum & $\rm O(^3P) + NO_3 \to NO_2 + O_2$ & $1.00\times10^{-11}$ & \cite{kossyi1992} \\
 \Rnum & $\rm O(^3P) + O(^3P) + O_2 \to 2O_2$ & $2.45\times10^{-31}T^{-0.63}$ & \cite{kossyi1992} \\
 \Rnum & $\rm O(^3P) + O(^3P) + N_2 \to O_2 + N_2$ & $2.76\times10^{-34}\exp(\frac{720}{T})$ & \cite{kossyi1992} \\
 \Rnum & $\rm O(^3P) + O(^1S) \to O(^3P) + O(^1D)$ & $5.00\times10^{-11}\exp(\frac{-301}{T})$ & \cite{kossyi1992} \\
 \Rnum & $\rm O(^1D) + O_2 \to O_2 + O(^3P)$ & $6.40\times10^{-12}\exp(\frac{67}{T})$ & \cite{kossyi1992} \\
 \Rnum & $\rm O(^1D) + O_2 \to O_2(b) + O(^3P)$ & $2.56\times10^{-11}\exp(\frac{67}{T})$ & \cite{kossyi1992} \\
 \Rnum & $\rm O(^1D) + O_3 \to 2O_2$ & $1.20\times10^{-10}$ & \cite{kossyi1992} \\
 \Rnum & $\rm O(^1D) + O_3 \to O_2 + 2O(^3P)$ & $1.20\times10^{-10}$ & \cite{kossyi1992} \\
 \Rnum & $\rm O(^1D) + N_2 \to N_2 + O(^3P)$ & $1.80\times10^{-11}\exp(\frac{107}{T})$ & \cite{kossyi1992} \\
 \Rnum & $\rm O(^1D) + NO \to O_2 + N(^4S)$ & $1.70\times10^{-10}$ & \cite{kossyi1992} \\
 \Rnum & $\rm O(^1D) + NO_2 \to NO + O_2$ & $3.00\times10^{-10}$ & \cite{gordillo-vazquez2008} \\
 \Rnum & $\rm O(^1D) + N_2O \to N_2 + O_2$ & $4.40\times10^{-11}$ & \cite{kossyi1992} \\
 \Rnum & $\rm O(^1D) + N_2O \to 2NO$ & $7.20\times10^{-11}$ & \cite{kossyi1992} \\
 \Rnum & $\rm O(^1S) + O_2 \to O_2 + O(^1D)$ & $1.33\times10^{-12}\exp(\frac{-850}{T})$ & \cite{kossyi1992} \\
 \Rnum & $\rm O(^1S) + O_2 \to O_2(A) + O(^3P)$ & $2.97\times10^{-12}\exp(\frac{-850}{T})$ & \cite{kossyi1992} \\
 \Rnum & $\rm O(^1S) + O_3 \to 2O_2$ & $2.90\times10^{-10}$ & \cite{kossyi1992} \\
 \Rnum & $\rm O(^1S) + O_3 \to O_2 + O(^3P) + O(^1D)$ & $2.90\times10^{-10}$ & \cite{kossyi1992} \\
 \Rnum & $\rm O(^1S) + NO \to NO + O(^3P)$ & $1.80\times10^{-10}$ & \cite{kossyi1992} \\
 \Rnum & $\rm O(^1S) + NO \to NO + O(^1D)$ & $3.20\times10^{-10}$ & \cite{kossyi1992} \\
 \Rnum & $\rm O(^1S) + N_2O \to N_2O + O(^3P)$ & $6.30\times10^{-12}$ & \cite{kossyi1992} \\
 \Rnum & $\rm O(^1S) + N_2O \to N_2O + O(^1D)$ & $3.10\times10^{-12}$ & \cite{kossyi1992} \\
 \Rnum & $\rm NO + O_3 \to NO_2 + O_2$ & $4.30\times10^{-12}\exp(\frac{-1560}{T})$ & \cite{kossyi1992} \\
 \Rnum & $\rm NO + NO + O_2 \to 2NO_2$ & $3.30\times10^{-39}\exp(\frac{530}{T})$ & \cite{atkinson2004} \\
 \Rnum & $\rm NO + NO_2 + N_2 \to N_2O_3 + N_2$ & $3.10\times10^{-34}(\frac{300}{T})^{7.7}$ & \cite{atkinson2004} \\
 \Rnum & $\rm NO + NO_2 + NO_3 \to NO + N_2O_5$ & $5.90\times10^{-29}(\frac{300}{T})^{1.27}$ & \cite{gordillo-vazquez2008} \\
 \Rnum & $\rm NO + NO_3 \to 2NO_2$ & $1.70\times10^{-11}$ & \cite{kossyi1992} \\
 \Rnum & $\rm NO_2 + O_3 \to NO_3 + O_2$ & $1.20\times10^{-13}\exp(\frac{-2450}{T})$ & \cite{kossyi1992} \\
 \Rnum & $\rm NO_2 + NO_2 + N_2 \to N_2O_4 + N_2$ & $1.40\times10^{-33}(\frac{300}{T})^{3.8}$ & \cite{atkinson2004} \\
 \Rnum & $\rm NO_2 + NO_3 + M \to N_2O_5 + M$ & $5.90\times10^{-29}(\frac{300}{T})^{1.27}$ & \cite{gordillo-vazquez2008} \\
 \Rnum & $\rm NO_2 + NO_3 \to NO + NO_2 + O_2$ & $2.30\times10^{-13}\exp(\frac{-1600}{T})$ & \cite{kossyi1992} \\
 \Rnum & $\rm NO_3 + NO_3 \to 2NO_2 + O_2$ & $5.00\times10^{-12}\exp(\frac{-3000}{T})$ & \cite{kossyi1992} \\
 \br
\end{tabular*}
\end{table*}

\section*{References}
\normalem
\bibliography{references}

\begin{thebibliography}{10}

\bibitem{nijdam2020a}
Sander Nijdam, Jannis Teunissen, and Ute Ebert.
\newblock The physics of streamer discharge phenomena.
\newblock {\em Plasma Sources Science and Technology}, 29(10):103001, November
  2020.

\bibitem{wang2020a}
Douyan Wang and Takao Namihira.
\newblock Nanosecond pulsed streamer discharges: {{II}}. {{Physics}}, discharge
  characterization and plasma processing.
\newblock {\em Plasma Sources Science and Technology}, 29(2):023001, February
  2020.

\bibitem{bruggeman2017}
Peter~J Bruggeman, Felipe Iza, and Ronny Brandenburg.
\newblock Foundations of atmospheric pressure non-equilibrium plasmas.
\newblock {\em Plasma Sources Science and Technology}, 26(12):123002, November
  2017.

\bibitem{kim2004}
Hyun-Ha Kim.
\newblock Nonthermal {{Plasma Processing}} for {{Air-Pollution Control}}: {{A
  Historical Review}}, {{Current Issues}}, and {{Future Prospects}}.
\newblock {\em Plasma Processes and Polymers}, 1(2):91--110, September 2004.

\bibitem{winands2006}
G.J.J. Winands, K.~Yan, A.J.M. Pemen, S.A. Nair, Z.~Liu, and E.J.M. Van~Heesch.
\newblock An {{Industrial Streamer Corona Plasma System}} for {{Gas Cleaning}}.
\newblock {\em IEEE Transactions on Plasma Science}, 34(5):2426--2433, October
  2006.

\bibitem{he2019}
Xianjun He, Yuxuan Zeng, Jialin Chen, Fushan Wang, Yunqing Fu, Fada Feng, and
  Haibao Huang.
\newblock Role of {{O}}{\textsubscript{3}} in the removal of {{HCHO}} using a
  {{DC}} streamer plasma.
\newblock {\em Journal of Physics D: Applied Physics}, 52(46):465203, November
  2019.

\bibitem{jiang2022}
Nan Jiang, Xiaoqi Kong, Xiaoling Lu, Bangfa Peng, Zhengyan Liu, Jie Li, Kefeng
  Shang, Na~Lu, and Yan Wu.
\newblock Promoting streamer propagation, active species generation and
  trichloroethylene degradation using a three-electrode nanosecond pulsed
  sliding {{DBD}} nanosecond plasma.
\newblock {\em Journal of Cleaner Production}, 332:129998, January 2022.

\bibitem{amirov1998}
Ravil~H. Amirov, Jae~O. Chae, Yuriy~N. Dessiaterik, Elena~A. Filimonova, and
  Mark~B. Zhelezniak.
\newblock Removal of {{NO}}{\textsubscript{{\it{x}}}} and
  {{SO}}{\textsubscript{{2}}} from {{Air Excited}} by {{Streamer Corona}}:
  {{Experimental Results}} and {{Modeling}}.
\newblock {\em Japanese Journal of Applied Physics}, 37(Part 1, No.
  6A):3521--3529, June 1998.

\bibitem{hazama2000}
Hisanao Hazama, Masanori Fujiwara, and Mitsumori Tanimoto.
\newblock Removal processes of nitric oxide along positive streamers observed
  by laser-induced fluorescence imaging spectroscopy.
\newblock {\em Chemical Physics Letters}, 323(5-6):542--548, June 2000.

\bibitem{kuroki2002}
T.~Kuroki, M.~Takahashi, M.~Okubo, and T.~Yamamoto.
\newblock Single-stage plasma-chemical process for particulates,
  {{NO}}{\textsubscript{{\it{x}}}}, and {{SO}}{\textsubscript{{\it{x}}}}
  simultaneous removal.
\newblock {\em IEEE Transactions on Industry Applications}, 38(5):1204--1209,
  September 2002.

\bibitem{sahni2006}
Mayank Sahni and Bruce~R. Locke.
\newblock Degradation of chemical warfare agent simulants using gas\textendash
  liquid pulsed streamer discharges.
\newblock {\em Journal of Hazardous Materials}, 137(2):1025--1034, September
  2006.

\bibitem{joshi2013}
Ravindra~P. Joshi and Selma~Mededovic Thagard.
\newblock Streamer-{{Like Electrical Discharges}} in {{Water}}: {{Part II}}.
  {{Environmental Applications}}.
\newblock {\em Plasma Chemistry and Plasma Processing}, 33(1):17--49, February
  2013.

\bibitem{bardos2010}
L.~B{\'a}rdos and H.~Bar{\'a}nkov{\'a}.
\newblock Cold atmospheric plasma: {{Sources}}, processes, and applications.
\newblock {\em Thin Solid Films}, 518(23):6705--6713, September 2010.

\bibitem{polonskyi2021}
O.~Polonskyi, T.~Hartig, J.~R. Uzarski, and M.~J. Gordon.
\newblock Precise localization of {{DBD}} plasma streamers using
  topographically patterned insulators for maskless structural and chemical
  modification of surfaces.
\newblock {\em Applied Physics Letters}, 119(21):211601, November 2021.

\bibitem{fridman2008}
Gregory Fridman, Gary Friedman, Alexander Gutsol, Anatoly~B. Shekhter,
  Victor~N. Vasilets, and Alexander Fridman.
\newblock Applied {{Plasma Medicine}}.
\newblock {\em Plasma Processes and Polymers}, 5(6):503--533, August 2008.

\bibitem{vonwoedtke2020}
Thomas {von Woedtke}, Steffen Emmert, Hans-Robert Metelmann, Stefan Rupf, and
  Klaus-Dieter Weltmann.
\newblock Perspectives on cold atmospheric plasma ({{CAP}}) applications in
  medicine.
\newblock {\em Physics of Plasmas}, 27(7):070601, July 2020.

\bibitem{starikovskaia2014}
S~M Starikovskaia.
\newblock Plasma-assisted ignition and combustion: Nanosecond discharges and
  development of kinetic mechanisms.
\newblock {\em Journal of Physics D: Applied Physics}, 47(35):353001, September
  2014.

\bibitem{popov2016}
N~A Popov.
\newblock Kinetics of plasma-assisted combustion: Effect of non-equilibrium
  excitation on the ignition and oxidation of combustible mixtures.
\newblock {\em Plasma Sources Science and Technology}, 25(4):043002, August
  2016.

\bibitem{kossyi1992}
I~A Kossyi, A~Yu Kostinsky, A~A Matveyev, and V~P Silakov.
\newblock Kinetic scheme of the non-equilibrium discharge in nitrogen-oxygen
  mixtures.
\newblock {\em Plasma Sources Science and Technology}, 1(3):207--220, August
  1992.

\bibitem{gordiets1995}
B.F. Gordiets, C.M. Ferreira, V.L. Guerra, J.M.A.H. Loureiro, J.~Nahorny,
  D.~Pagnon, M.~Touzeau, and M.~Vialle.
\newblock Kinetic model of a low-pressure
  {{N}}\textsubscript{2}-{{O}}\textsubscript{2} flowing glow discharge.
\newblock {\em IEEE Transactions on Plasma Science}, 23(4):750--768, August
  1995.

\bibitem{aleksandrov1999}
N~L Aleksandrov and E~M Bazelyan.
\newblock Ionization processes in spark discharge plasmas.
\newblock {\em Plasma Sources Science and Technology}, 8(2):285--294, May 1999.

\bibitem{fresnet2002}
F~Fresnet, G~Baravian, L~Magne, S~Pasquiers, C~Postel, V~Puech, and A~Rousseau.
\newblock Influence of water on {{NO}} removal by pulsed discharge in
  {{N}}{\textsubscript{2}}/{{H}}{\textsubscript{2}}{{O}}/{{NO}} mixtures.
\newblock {\em Plasma Sources Science and Technology}, 11(2):152--160, May
  2002.

\bibitem{tochikubo2002}
Fumiyoshi Tochikubo and Hideyuki Arai.
\newblock Numerical {{Simulation}} of {{Streamer Propagation}} and {{Radical
  Reactions}} in {{Positive Corona Discharge}} in {N}\textsubscript{2}/{NO} and
  {N}\textsubscript{2}/{O}\textsubscript{2}/{NO}.
\newblock {\em Japanese Journal of Applied Physics}, 41(Part 1, No.
  2A):844--852, February 2002.

\bibitem{atkinson2004}
R.~Atkinson, D.~L. Baulch, R.~A. Cox, J.~N. Crowley, R.~F. Hampson, R.~G.
  Hynes, M.~E. Jenkin, M.~J. Rossi, and J.~Troe.
\newblock Evaluated kinetic and photochemical data for atmospheric chemistry:
  {{Volume I}} - gas phase reactions of {O}\textsubscript{\it{x}},
  {HO}\textsubscript{\it{x}}, {NO}\textsubscript{\it{x}} and
  {SO}\textsubscript{\it{x}} species.
\newblock {\em Atmospheric Chemistry and Physics}, 4(6):1461--1738, September
  2004.

\bibitem{florescumitchell2006}
A~Florescumitchell and J~Mitchell.
\newblock Dissociative recombination.
\newblock {\em Physics Reports}, 430(5-6):277--374, August 2006.

\bibitem{gordillo-vazquez2008}
F~J {Gordillo-V{\'a}zquez}.
\newblock Air plasma kinetics under the influence of sprites.
\newblock {\em Journal of Physics D: Applied Physics}, 41(23):234016, December
  2008.

\bibitem{popov2011}
N~A Popov.
\newblock Fast gas heating in a nitrogen\textendash oxygen discharge plasma:
  {{I}}. {{Kinetic}} mechanism.
\newblock {\em Journal of Physics D: Applied Physics}, 44(28):285201, July
  2011.

\bibitem{pancheshnyi2013}
Sergey Pancheshnyi.
\newblock Effective ionization rate in nitrogen\textendash oxygen mixtures.
\newblock {\em Journal of Physics D: Applied Physics}, 46(15):155201, April
  2013.

\bibitem{ono2020}
Ryo Ono and Atsushi Komuro.
\newblock Generation of the single-filament pulsed positive streamer discharge
  in atmospheric-pressure air and its comparison with two-dimensional
  simulation.
\newblock {\em Journal of Physics D: Applied Physics}, 53(3):035202, January
  2020.

\bibitem{bouwman2022}
Dennis Bouwman, Jannis Teunissen, and Ute Ebert.
\newblock {{3D}} particle simulations of positive air\textendash methane
  streamers for combustion.
\newblock {\em Plasma Sources Science and Technology}, 31(4):045023, April
  2022.

\bibitem{komuro2015}
Atsushi Komuro, Kazunori Takahashi, and Akira Ando.
\newblock Numerical simulation for the production of chemically active species
  in primary and secondary streamers in atmospheric-pressure dry air.
\newblock {\em Journal of Physics D: Applied Physics}, 48(21):215203, June
  2015.

\bibitem{naidis1997a}
G~V Naidis.
\newblock Modelling of plasma chemical processes in pulsed corona discharges.
\newblock {\em Journal of Physics D: Applied Physics}, 30(8):1214--1218, April
  1997.

\bibitem{naidis2012}
G~V Naidis.
\newblock Efficiency of generation of chemically active species by pulsed
  corona discharges.
\newblock {\em Plasma Sources Science and Technology}, 21(4):042001, August
  2012.

\bibitem{vanveldhuizen1996}
E.~M. {van Veldhuizen}, W.~R. Rutgers, and V.~A. Bityurin.
\newblock Energy efficiency of {{NO}} removal by pulsed corona discharges.
\newblock {\em Plasma Chemistry and Plasma Processing}, 16(2):227--247, June
  1996.

\bibitem{namihira2000}
T.~Namihira, S.~Tsukamoto, Douyan Wang, S.~Katsuki, R.~Hackam, H.~Akiyama,
  Y.~Uchida, and M.~Koike.
\newblock Improvement of {NO}{\textsubscript{{\it{x}}}} removal efficiency
  using short-width pulsed power.
\newblock {\em IEEE Transactions on Plasma Science}, 28(2):434--442, April
  2000.

\bibitem{matsumoto2010}
Takao Matsumoto, Douyan Wang, Takao Namihira, and Hidenori Akiyama.
\newblock Energy {{Efficiency Improvement}} of {{Nitric Oxide Treatment Using
  Nanosecond Pulsed Discharge}}.
\newblock {\em IEEE Transactions on Plasma Science}, 38(10):2639--2643, October
  2010.

\bibitem{teunissen2017}
Jannis Teunissen and Ute Ebert.
\newblock Simulating streamer discharges in {{3D}} with the parallel adaptive
  {{Afivo}} framework.
\newblock {\em Journal of Physics D: Applied Physics}, 50(47):474001, November
  2017.

\bibitem{zheleznyak1982}
M~B {Zheleznyak}, A~Kh {Mnatsakanyan}, and S~V {Sizykh}.
\newblock Photoionization of nitrogen and oxygen mixtures by radiation from a
  gas discharge.
\newblock {\em High Temperature}, 20(3):357--362, November 1982.

\bibitem{luque2007}
Alejandro Luque, Ute Ebert, Carolynne Montijn, and Willem Hundsdorfer.
\newblock Photoionization in negative streamers: {{Fast}} computations and two
  propagation modes.
\newblock {\em Applied Physics Letters}, 90(8):081501, February 2007.

\bibitem{bourdon2007}
A~Bourdon, V~P Pasko, N~Y Liu, S~C{\'e}lestin, P~S{\'e}gur, and E~Marode.
\newblock Efficient models for photoionization produced by non-thermal gas
  discharges in air based on radiative transfer and the {{Helmholtz}}
  equations.
\newblock {\em Plasma Sources Science and Technology}, 16(3):656--678, August
  2007.

\bibitem{guo2022d}
Baohong Guo, Xiaoran Li, Ute Ebert, and Jannis Teunissen.
\newblock A computational study of accelerating, steady and fading negative
  streamers in ambient air.
\newblock {\em Plasma Sources Science and Technology}, 31(9):095011, September
  2022.

\bibitem{komuro2012}
Atsushi Komuro, Ryo Ono, and Tetsuji Oda.
\newblock Numerical simulation for production of {{O}} and {{N}} radicals in an
  atmospheric-pressure streamer discharge.
\newblock {\em Journal of Physics D: Applied Physics}, 45(26):265201, July
  2012.

\bibitem{li2022a}
Xiaoran Li, Baohong Guo, Anbang Sun, Ute Ebert, and Jannis Teunissen.
\newblock A computational study of steady and stagnating positive streamers in
  {{N}} {\textsubscript{2}} \textendash{{O}} {\textsubscript{2}} mixtures.
\newblock {\em Plasma Sources Science and Technology}, 31(6):065011, June 2022.

\bibitem{teunissen2018}
Jannis Teunissen and Ute Ebert.
\newblock Afivo: {{A}} framework for quadtree/octree {{AMR}} with shared-memory
  parallelization and geometric multigrid methods.
\newblock {\em Computer Physics Communications}, 233:156--166, December 2018.

\bibitem{teunissen2023}
Jannis Teunissen and Francesca Schiavello.
\newblock Geometric multigrid method for solving {{Poisson}}'s equation on
  octree grids with irregular boundaries.
\newblock {\em Computer Physics Communications}, 286:108665, May 2023.

\bibitem{hagelaar2005}
G~J~M Hagelaar and L~C Pitchford.
\newblock Solving the {{Boltzmann}} equation to obtain electron transport
  coefficients and rate coefficients for fluid models.
\newblock {\em Plasma Sources Science and Technology}, 14(4):722--733, November
  2005.

\bibitem{phelps_database}
Phelps database ({N}\textsubscript{2}, {O}\textsubscript{2})
  \url{www.lxcat.net} (retrieved 8 June 2022).

\bibitem{lawton1978}
S.~A. Lawton and A.~V. Phelps.
\newblock Excitation of the {b}\textsuperscript{1}{$\Sigma$}$_g^+$ state of
  {O}\textsubscript{2} by low energy electrons.
\newblock {\em The Journal of Chemical Physics}, 69(3):1055, August 1978.

\bibitem{phelps1985}
A.~V. Phelps and L.~C. Pitchford.
\newblock Anisotropic scattering of electrons by {{N}}{\textsubscript{{2}}} and
  its effect on electron transport.
\newblock {\em Physical Review A}, 31(5):2932--2949, May 1985.

\bibitem{li2021a}
Xiaoran Li, Siebe Dijcks, Sander Nijdam, Anbang Sun, Ute Ebert, and Jannis
  Teunissen.
\newblock Comparing simulations and experiments of positive streamers in air:
  Steps toward model validation.
\newblock {\em Plasma Sources Science and Technology}, 30(9):095002, September
  2021.

\bibitem{nijdam2011}
S~Nijdam, G~Wormeester, E~M {van Veldhuizen}, and U~Ebert.
\newblock Probing background ionization: Positive streamers with varying pulse
  repetition rate and with a radioactive admixture.
\newblock {\em Journal of Physics D: Applied Physics}, 44(45):455201, November
  2011.

\bibitem{francisco2021e}
Hani Francisco, Jannis Teunissen, Behnaz Bagheri, and Ute Ebert.
\newblock Simulations of positive streamers in air in different electric
  fields: Steady motion of solitary streamer heads and the stability field.
\newblock {\em Plasma Sources Science and Technology}, 30(11):115007, November
  2021.

\bibitem{luque2008}
Alejandro Luque, Valeria Ratushnaya, and Ute Ebert.
\newblock Positive and negative streamers in ambient air: Modelling evolution
  and velocities.
\newblock {\em Journal of Physics D: Applied Physics}, 41(23):234005, December
  2008.

\bibitem{starikovskiy2020}
A~Yu Starikovskiy and N~L Aleksandrov.
\newblock How pulse polarity and photoionization control streamer discharge
  development in long air gaps.
\newblock {\em Plasma Sources Science and Technology}, 29(7):075004, July 2020.

\bibitem{vanheesch2008}
E~J~M {van Heesch}, G~J~J Winands, and A~J~M Pemen.
\newblock Evaluation of pulsed streamer corona experiments to determine the
  {{O}}{\textsuperscript{*}} radical yield.
\newblock {\em Journal of Physics D: Applied Physics}, 41(23):234015, December
  2008.

\bibitem{ono2011}
Ryo Ono, Yusuke Nakagawa, and Tetsuji Oda.
\newblock Effect of pulse width on the production of radicals and excited
  species in a pulsed positive corona discharge.
\newblock {\em Journal of Physics D: Applied Physics}, 44(48):485201, December
  2011.

\bibitem{winands2006a}
G~J~J Winands, Z~Liu, A~J~M Pemen, E~J~M van Heesch, K~Yan, and E~M van
  Veldhuizen.
\newblock Temporal development and chemical efficiency of positive streamers in
  a large scale wire-plate reactor as a function of voltage waveform
  parameters.
\newblock {\em Journal of Physics D: Applied Physics}, 39(14):3010--3017, July
  2006.

\bibitem{wang2010}
Douyan Wang, Takao Matsumoto, Takao Namihira, and Hidenori Akiyama.
\newblock Development of {{Higher Yield Ozonizer Based}} on {{Nano-Seconds
  Pulsed Discharge}}.
\newblock {\em Journal of Advanced Oxidation Technologies}, 13(1):71--78,
  January 2010.

\bibitem{bagheri2020}
Behnaz Bagheri, Jannis Teunissen, and Ute Ebert.
\newblock Simulation of positive streamers in {{CO}}{\textsubscript{2}} and in
  air: The role of photoionization or other electron sources.
\newblock {\em Plasma Sources Science and Technology}, 29(12):125021, December
  2020.

\bibitem{simek2014}
M~{\v S}imek.
\newblock Optical diagnostics of streamer discharges in atmospheric gases.
\newblock {\em Journal of Physics D: Applied Physics}, 47(46):463001, November
  2014.

\bibitem{komuro2015a}
Atsushi Komuro, Kazunori Takahashi, and Akira Ando.
\newblock Vibration-to-translation energy transfer in atmospheric-pressure
  streamer discharge in dry and humid air.
\newblock {\em Plasma Sources Science and Technology}, 24(5):055020, September
  2015.

\bibitem{guthrie1991}
J.~A. Guthrie, R.~C. Chaney, and A.~J. Cunningham.
\newblock Temperature dependencies of ternary ion\textendash molecule
  association reactions yielding {N}$_3^+$, {N}$_4^+$, and {CO}$_2^+$.
\newblock {\em The Journal of Chemical Physics}, 95(2):930--936, July 1991.

\bibitem{matzing1991}
H.~M{\"a}tzing.
\newblock Chemical {{Kinetics}} of {{Flue Gas Cleaning}} by {{Irradiation}}
  with {{Electrons}}.
\newblock In {\em Advances in {{Chemical Physics}}}, pages 315--402. {John
  Wiley \& Sons, Ltd}, 1991.

\bibitem{sutherland1975}
C.D. Sutherland and J.~Zinn.
\newblock Chemistry computations for irradiated hot air.
\newblock Technical Report LA--6055-MS, 4140364, August 1975.

\bibitem{DNA1973}
Defense Nuclear Agency Reaction Rate Handbook, Second Edition, Revision No. 3,
  DNA 1948H (September 1973).

\end{thebibliography}

\end{document}